\documentclass[%
 reprint,
 amsmath,amssymb,
 aps,
floatfix
]{revtex4-2}

\usepackage{graphicx}
\usepackage{dcolumn}
\usepackage{bm}
\usepackage{lineno}
\usepackage{float}
\usepackage{xcolor}
\usepackage{appendix}
\usepackage{hyperref}
\usepackage{lineno}

\definecolor{mygreen}{rgb}{0., 0.5, 0.}

\newcommand{\crittemp}{T_I^c}
\newcommand{\critmu}{\mu_I^c}

\newcommand{\Ob}{\mathcal{O}}

\newcommand{\muI}{\mu_I}
\newcommand{\muB}{\mu_B}

\newcommand{\M}{\mathcal{M}}

\newcommand{\C}{\mathcal{C}}

\newcommand{\modmuI}{\left|\muI\right|}

\newcommand{\LeeYangzeroisospin}{\muI^{\text{LY}}}

\newcommand{\ROCresum}{\rho_{\text{R}}}
\newcommand{\ROCTresum}{\mu_I^{\rho}}

\newcommand{\nearestLeeYangzeroisospin}{\muI^0}
\newcommand{\nearestLeeYangzeroisospinreal}{\text{Re}\left[\nearestLeeYangzeroisospin\right]}
\newcommand{\nearestLeeYangzeroisospinimag}{\text{Im}\left[\nearestLeeYangzeroisospin\right]}

\newcommand{\quickLYRe}{\mu_{I,r}^0}
\newcommand{\quickLYIm}{\mu_{I,i}^0}

\newcommand{\quickLYReT}{\mu_{I,r}^{T}}
\newcommand{\quickLYImT}{\mu_{I,i}^{T}}

\newcommand{\LA}{\left \langle}
\newcommand{\RA}{\right \rangle}

\newcommand{\Nt}{\text{N}_{\tau}}

\newcommand{\hs}{\hspace{.6cm}}

\newcommand{\partitionfunction}{\mathcal{Z}}

\newcommand{\susc}{\mathcal{X}}

\newcommand{\TaylorpressuretoN}{\frac{\Delta P_N^{\text{T}}}{T^4}}

\newcommand{\ResummedpressuretoN}{\frac{\Delta P_N^{\text{R}}}{T^4}}

\usepackage{float}
\usepackage{calc}
\newlength{\depthofsumsign}
\newcommand{\nsum}[1][1.2]{
    \mathop{%
        \raisebox
            {-#1\depthofsumsign+1\depthofsumsign}
            {\scalebox
                {#1}
                {$\displaystyle\sum$}%
            }
    }
}

\begin{document}

\preprint{APS/123-QED}

\title{Estimates of Lee-Yang zeros and possible critical point on pion condensate boundary in QCD isospin phase diagram using unbiased exponential resummation on Lattice}

\author{Sabarnya Mitra}
\email{smitra@physik.uni-bielefeld.de}
\affiliation{Centre for High Energy Physics, Indian Institute of Science Bengaluru 560012, India}
 \affiliation{Fakult\"at f\"ur Physik, Universit\"at Bielefeld, D-33615 Bielefeld,
Germany}

\date{\today}

\begin{abstract}

  Without invoking any cumulant determination at the input level, we present here the first calculations of direct estimates of the Lee-Yang zeros of QCD partition function in (2+1)-flavor QCD. These zeros are obtained in complex isospin chemical potential $\muI$ plane using the unbiased exponential resummation formalism on $N_\tau=8$ lattices and with physical quark masses. For different temperatures, we illustrate the stability of the zeros closest to the origin from which, we subsequently procure the radius of convergence estimates. From the temperature-dependence study of the real and imaginary parts of these zeros, we try estimating one of the possible critical points forming the second order pion condensate critical line in the isospin phase diagram. Further, we compare these resummed estimates with the corresponding Mercer-Roberts estimates of the subsequent Taylor series expansions of the first three partition function cumulants. We also outline comparisons between resummed and Taylor series results of these cumulants for real and imaginary values of $\muI$ and highlight the behavior of different expansion orders within and beyond the obtained resummed estimates of radius of convergence. We also re-establish that this resummed radius of convergence can efficiently capture the onset of overlap problem for finite real $\muI$ simulations.
 
\end{abstract}

\maketitle

\section{Introduction}
\label{sec:Intro}

The thermodynamics of Quantum Chromodynamics (QCD) under extreme conditions of temperature and baryon densities governs a wide spectrum of phenomena ranging from the creation of quark–gluon plasma \cite{Shuryak1978ij,ALICE2008ngc} in relativistic heavy-ion collision experiments \cite{Gyulassy2004RHIC,PHENIX2004RHIC,Aad2008ATLAS,Busza2018RHIC,Ratti2018RHIC} to the dense cores of neutron stars \cite{Baym1971star,Shuryak1980star,Baym2017star} and pion condensate formation \cite{Barshay1973pioncond}, apart from enlightening the early stages \cite{Kolb1990uni} of the universe and its subsequent evolution.  While the QCD phase diagram \cite{Halasz1998qcd_pha_dia,Karsch2001QCD,HotQCD2014QCD} and its thermodynamics in the plane of temperature $T$ \cite{HotQCD2018crossover,Steinbrecher2018crossover,Borsanyi2020crossover} and baryon chemical potential $\muB$ \cite{Karsch1988fin_den_qcd,Karsch1999fin_den_qcd,deForcrand2009fin_den_qcd,Fu2019fin_den_qcd} is of prime relevance and utmost importance both from theoretical \cite{Fodor2001crit,Karsch2001crit,Gavai2004crit} and experimental \cite{Gyulassy2004RHIC,PHENIX2004RHIC,Aad2008ATLAS} perspectives, the notorious sign problem \cite{Barbour1986jf,Fodor2001au,Pan2022fgf} by complexifying Monte-Carlo integral measure and precluding standard importance sampling techniques, hinders any possible direct effort of mapping this phase diagram and drawing conclusive evidences from first-principle lattice QCD calculations. Despite numerous approaches \cite{deForcrand2003IMag,Gavai2004Taylor,Sexty2013ComLange,Alexandru2015thimbles,Mondal2021exp_res,Bollweg2022Pade,Mitra2023unb_exp_res} of circumventing the problem and uncovering the phase diagram, most of it is still in the dark and requires further convincing results. This is one of the very motivations in exploiting an alternative avenue in the direction of the isospin chemical potential $\mu_{I}=(\mu_{u}-\mu_{d})$ with up and down quark chemical potentials $\mu_u, \,\,\mu_d$ which unlike $\muB$ introduces an asymmetry between up and down quark densities without introducing a sign problem. This inspires profound interest in using direct lattice QCD simulations for the purpose of exploring the resulting $T$–$\mu_{I}$ isospin phase diagram of QCD \cite{Son2000xc,deForcrand2007isophdia,Andersen2023ofv,Brandt2017isophdia,Brandt2019isophdia}. Besides manifesting the familiar hadronic and quark-gluon plasma phases respectively below and above their smooth crossover line \cite{HotQCD2018crossover,Borsanyi2020crossover} around $T \approx 157$ MeV at zero $\muI$, the various state-of-the-art model studies also predict two other principal phases and regimes in this isospin phase diagram for low $T$: (i) a Bose–Einstein condensate (BEC) of charged pions for $\mu_{I}>m_{\pi}$, characterized by a second-order transition of $O(2)$ universality class \cite{Brandt2017isophdia}, and (ii) a smooth crossover from the pionic BEC state into a Bardeen–Cooper–Schrieffer (BCS) color-superconducting phase at large $\mu_{I}$ \cite{Son2000xc}, where quark Cooper pairs form in the pseudo-scalar channel \cite{Yokota2023osv}. While the latter is explained from perturbative arguments \cite{Reinosa2015oua}, the former is well-provided by the chiral perturbation theory \cite{Scherer2002tk} predictions, which holds reliably in the low-energy regime of QCD. Despite significant progress using functional methods \cite{Fu2019fin_den_qcd}, chiral effective theories \cite{Son2000xc} and both canonical \cite{Alexandru2005la} and grand-canonical lattice approaches, quantitative details regarding the nature and curvature of transition lines separating the relevant phases for different values of $T,\muI$ remain elusive and demands further extensive searches. Similarly, the perturbative predictions concerning the possible emergence of a first-order transition at asymptotically large values of $\mu_{I}$ accompanied by decoupling of the gluon sector also remain under active investigation.

In this work, the central focus is on the pion phase boundary that separates the pion condensate phase from the hadronic non-condensate phase. At low temperatures and small values of $\muI$, the phase diagram is characterized by the formation of a pion condensate. As $\muI$ increases, the energy required to create charged pions reduces due to their coupling to the isospin chemical potential $\muI$. When $\muI$ reaches the critical value $\mu_{I,c}$, it becomes energetically favorable for charged pions to condense into a Bose-Einstein-like ground state. With this value of $\mu_{I,c}$ depending on the initial choice of convention of the theory, current studies in the literature toggle between $\mu_{I,c}=m_\pi$ \cite{Son2000xc} and  $m_\pi/2$ \cite{Brandt2017isophdia} for $T=0$ in the phase diagram, where this transition into the pion condensed phase across the pion phase boundary line is a second-order phase transition characterized by a nonzero expectation value of the charged pion field $\LA \pi^{\pm}\RA$ and the spontaneous breaking of the residual $U(1)$ isospin symmetry, accompanied by the prescence of a Goldstone mode. This transition occurs primarily at low temperatures, where there are less thermal fluctuations, which are insufficient to disrupt the coherence of the condensate. While chiral perturbation theory ($\chi$PT) reliable in this regime of small $T,\muI$ predicts a monotonic slope of this critical line with increasing value of critical $\muI$ for rising $T$, recentmost studies \cite{Brandt2017isophdia} have found this line to remain vertical in the $T$-$\muI$ plane upto $T\approx140$ MeV starting from $\mu_{I,c}=m_\pi/2$ ($m_\pi$ here and also in \cite{Son2000xc,Andersen2023ofv}) at $T=0$, before the thermal effect starts dominating for higher temperatures and thereby melting the condensate close to and beyond the crossover temperature $T \approx 157$ MeV. Further future studies are required for a more firm statement regarding the thermal nature of this pion condensate critical line.

In the present work, we leverage the recently-developed method of unbiased exponential resummation \cite{Mitra2023unb_exp_res,Mitra2023thesis} for probing the analytic structure of the isospin partition function in the complex $\mu_{I}$ plane without introducing any form of pionic regulator in the working QCD action here, where the central goal is to have an understanding of the QCD isospin phase diagram in the low $T$, small $\muI$ regime from the perspective of Lee-Yang zeros and related non-analytic structure of the isospin partition function. By obtaining and locating the Lee–Yang zeros of the partition function in its polynomial form in the complex $\muI$ plane using Newton-Raphson method, we procure a non-perturbative resummed estimate of the radius of convergence, by determining the closest Lee-Yang zero from the origin and evaluating its distance from the origin. As outlined in detail in this paper, we also verify the stability of these zeros which as we will observe, subsequently yields encouraging indications. This is promising, as these zeros and the associated radius of convergence offer valuable insights regarding the proximity of possible phase transition in $\muI$ although strictly speaking, one needs to investigate these zeros in the thermodynamic limit for confirming possible phase transition or crossover. With information from the study of the temperature dependence of the radius of convergence estimates and the same for the imaginary parts of the nearest complex Lee-Yang zeros, we procure a quantitative measurement of the estimate of a true critical point in the isospin phase diagram. We show and discuss in length of Section \ref{sec:Radius of convergence} of this paper, that this critical point obtained for the first time in this manner, happen to lie on the second order pion phase boundary critical line within the respective errorbars for the respective $T$ and $\muI$ values. This agreement as we show in the paper, is consistent with the starting convention of this work which otherwise influences the critical $\mu_{I,c}$ marking the onset of the pion phase. 
Although this has no implication on the order and universality class of this phase transition, observing the critical point on the pion phase boundary from a direct Lee-Yang zero analysis at this low temperature, is a novel feature of this work and additionally this is achieved without relying on the knowledge of the partition function cumulants which is unlike previous works \cite{Dimopoulos2021vrk,Schmidt2022ogw,Clarke2024seq,Clarke2024ugt} in prevailing lattice QCD literature on Lee-Yang zero \cite{LeeYang1952} analysis and determination.  
Through this work and its pertinent study, we strive towards attaining a quantitative characterization of the QCD isospin phase diagram. This attempt will not only refine our existent knowledge of pion condensation at finite isospin density from the new unbiased resummation approach, but also pave way in the future for similar studies and related insights for the more experimentally relevant finite baryon densities. Besides comparing this resummed estimate of the radius of convergence with the corresponding Mercer-Roberts \cite{Mercer1990,Pasztor2019mercer,Vovchenko2017cluster} estimates of the same obtained from the Taylor expansion series, we also illustrate comparison between the resummed and the Taylor series results of various order cumulants and outline the behaviour of different orders within and beyond this unbiased resummed estimate of the radius of convergence for several working temperatures presented in this paper. This has been shown here using an eighth order Taylor expansion around $\muI=0$ for the first three cumulants of the partition function \cite{Gavai2003tay,Ejiri2003tay,Miao2008tay,Karsch2010tay}. To summarize in brief, the flow of the work presented here, is as follows: 1) computing nearest Lee–Yang zeros on lattices with physical quark masses using Newton–Raphson algorithm in complex $\muI$ plane, 2) tracking the temperature dependence of their real and imaginary parts to gauge the possible critical point, and 3) benchmarking the resummed convergence radii against the stable Mercer–Roberts estimators derived from high-order Taylor coefficients. We also briefly outline the onset of the overlap problem across the obtained resummed radius towards the end of the paper.

The paper is therefore organized as follows : In Sec. \ref{sec:notations}, we motivate and review the formalism of finite-density Taylor series expansions by introducing the QCD partition function. We also briefly outline in this section, the unbiased exponential resummation method and the Newton–Raphson procedure, both of which forms a central part of this work and has been implemented in length in this paper. Although well-known, we brief these methods and introduce the basic notations in this section, which we extensively use here to estimate the respective Lee-Yang zeros.  Sec. \ref{sec:Lattice setup} details our lattice setup and the related simulation parameters of the results presented in this paper. We also illustrate in this section, the relevant quark and pion quantum numbers thus influencing the onset point of the pion condensate in the phase diagram. 
From Sec. \ref{sec:muI} onwards, we demonstrate the central results of this work in $\muI$ with vivid arguments. Instrumental to the work here, this core section is devoted to the computation and analysis of Lee–Yang zeros in the complex $\muI$ plane described in \ref{subsec:LY zeros} including a vivid discussion in \ref{subsec:stability} on their stability and extent of reliability for extracting meaningful results and subsequent conclusions. As we will observe, the Lee-Yang zero situated nearest to the origin yields maximal relevance since it offers the estimate of the resummed radius of convergence, which is subsequently mapped to possible determination of critical point and phase boundary. Sec.\ref{sec:Real and imaginary} discusses the temperature-dependence of the real and imaginary parts of this closest Lee-Yang zero for different working temperatures. From these data, the linear extrapolation towards a possible critical point at some critical $T$ and $\muI$ is performed, which is clearly mentioned and presented in \ref{subsec:estimates from extrapolation of zeros for lower T} of Sec.\ref{sec:Radius of convergence}, which also showcases the radius of convergence estimates for various temperature. Besides demonstrating the measure of the critical point and its location on the pion phase critical line, we also illustrate comparisons between this resummed estimate of radius of convergence and the Mercer-Roberts measures of the same, as shown in \ref{subsec: Ratio and MR estimates}. The latter two estimates are obtained from Taylor series results. These results have been obtained using the recent HotQCD data with more improved statistics for $N_\tau=8$ lattices \cite{Bollweg2022coefficients}. We also extend this comparative study to the level of individual cumulants of partition function in Sec.\,\ref{subsec: comparing Cumulants}, where as we will observe, there is order-by-order disagreement beyond this resummed radius of convergence. Following this, we discuss the overlap problem in Sec. \ref{sec:Overlap problem} and conclude in Sec. \ref{sec:Conclusions} with an outlook on possible future extensions. In the appendix part of the paper, we discuss 
the choice and reasoning of our starting isospin convention and the subsequent relevant quantum numbers of $u,d$ quarks and pion in \ref{appendix sec: fixing the convention}. Throughout this paper, we have used relativistic units and unit Boltzmann constant. Except Sec.\,\ref{sec:Taylor and Expo}, we use $\muB$ and $\muI$ to denote $\muB/T$ and $\muI/T$ in this paper.

\section{Essential Notations and formulae}
\label{sec:notations}

 In this section, we introduce some basic notations and revisit some well-known formulae which we have used here in this paper.  
 
\subsection{Taylor Expansion and Unbiased Exponential Resummation} \label{sec:Taylor and Expo}

 For (2+1)-flavor QCD with gluons described by a Symanzik-improved\,\cite{Symanzik1983gau} gauge action $S_g$ and quarks by the  Highly Improved Staggered Quark (HISQ) action\,\cite{Follana2006rc,Bazavov2010hisq,Bazavov2011nk}, the grand-canonical partition function $\partitionfunction(\mu,T)$ in thermodynamic limit for a temperature $T$ and chemical potential $\mu$ is represented by 

\begin{equation}
    \hspace{-3mm} \partitionfunction \left(T, \mu\right) = \int \mathcal{D}U \,e^{-S_g\left[U\left(T\right)\right]} \, \left[\det \M\left(\mu,T,U\right)\right]^{1/4}
    \label{eq:QCD partition function}
\end{equation}
with Euclidean gauge action $S_g$ and fermion matrix $\M$. This partition function is important for constructing the different order $\mu$ cumulants useful for knowing finite density QCD thermodynamics. 
The finite $\mu$ Taylor Expansion of the QCD excess pressure  yields an even series in $\mu$ owing to the particle-antiparticle symmetry of QCD. Upto $\Ob(\mu^N)$, this is given by

\begin{equation}
    \TaylorpressuretoN = \nsum_{n=1}^{N/2} \frac{\susc_{2n}}{(2n)!} \left(\frac{\mu}{T}\right)^{2n}
    \label{eq:QCD taylor expansion}
\end{equation}
where in Eqn.\eqref{eq:QCD taylor expansion}, $\Delta P_N^{\text{T}}$ is the Taylor estimate of excess pressure $\Delta P(\mu)$ defined as $\Delta P(\mu)=P(\mu)-P(0)$, and the Taylor coefficients $c_{2n}=\chi_{2n}/(2n)!$

Unlike the Taylor expansion, the unbiased exponential resummation starts its approximation from the level of $\partitionfunction$, and harnesses an estimate of $\Delta P(\mu)$ comprising a series to all orders in $\mu$. This is given by   

\begin{align}
    &\ResummedpressuretoN = \frac{1}{VT^3} \ln \left[\frac{\partitionfunction (\mu)}{\partitionfunction (0)}\right]_{\text{R}}, \hs \text{where}\notag \\ 
    &\left[\frac{\partitionfunction (\mu)}{\partitionfunction (0)}\right]_{\text{R}} = \LA e^{A_N\left(\mu\right)}\RA \hs \text{with}\notag \\  
    &A_N\left(\mu\right) = \nsum_{n=1}^N \,\,\,\frac{\C_n}{n!}\,\left(\frac{\mu}{T}\right)^n 
    \label{eq:QCD exponential resummation}
\end{align}
These coefficients $C_n$ in above Eqn.\eqref{eq:QCD exponential resummation} are constructed from different linear combinations of the various $n$-point correlation functions $D_n$. An elaborate study of this, with mathematical details is given in \cite{Mitra2023unb_exp_res,Mitra2023thesis}, whereas unbiased estimates and their importance are laid in \cite{Mitra2022cum_exp,Mitra2022unb_exp_res,Mitra2022Bonn_unb_exp_res}. We have seen for a given order $n$, this formalism yields the exact identical Taylor coefficients of each order $\leq n$, as well as provide non-zero estimates for all the higher orders. We will extensively use this method of resummation for all our work presented in this paper.   

\subsection{Newton-Raphson method}
\label{subsec:Newton-Raphson}

By using the unbiased exponential resummation to approximate the partition function $\mathcal{Z}(\mu)$ from its path integral form, we explicitly make use of the Newton-Raphson method in this paper to numerically evaluate the Lee-Yang zeros of this $\partitionfunction$, as a function of $\muI$. Starting from an initial guess $\mu_0$, one converges towards the actual analytic root of the function $f(\mu) \equiv (\partitionfunction(\mu)$ in our work) through several iterations following the recursive relation

\begin{equation}
    \mu_{n+1}=\mu_n-\frac{f(\mu_n)}{f^{'}(\mu_n)}
    \label{eq:Newton-Raphson}
\end{equation}
Here, $f^{'}(\mu_n)$ indicates the first derivative of $f$ {\it w.r.t} $\mu$ at $\mu=\mu_n$, the value of the root at $n$th iteration. The algorithm is designed so that these iterations go on till a maximum allowed value $N_{U}$ and values of $\mu_{n+1}$ are generated sequentially, unless $\left|\mu_{n+1}-\mu_n\right| \leq \epsilon$, where $\epsilon$ is the tolerance limit value of the given Newton-Raphson method. The values of the initial guess $\mu_0$ and the tolerance $\epsilon$ determines the pace of convergence of this method \cite{NewtonRaphson}. One keeps the upper limit $N_U$ of these iterations sufficiently high so that the target tolerance is reached, ensuring the algorithm never fails to attain the root value for given $\mu_0$ and $\epsilon$, for $1 \leq n \leq N_U$ as per Eqn.\eqref{eq:Newton-Raphson}.

\section{Setup of lattice and simulations}
\label{sec:Lattice setup}
For all the analysis presented in this paper, we have used the data generated by the HotQCD collaboration for its ongoing Taylor expansion calculations of the finite density QCD Equation of State (EoS), chiral crossover temperature and conserved charge cumulants at finite density~\cite{Bazavov:2017dus,HotQCD:2018pds,Bollweg:2021vqf,Bollweg:2022rps,Bollweg:2022fqq}. A detailed overview and description of the gauge ensembles and scale setting can be found in Ref.~\cite{Bollweg:2021vqf}. In this HotQCD data, (2+1)-flavor gauge configurations are generated in the temperature range $T \in [125:176]$~MeV using a Symanzik-improved gauge action and Highly Improved Staggered Quark (HISQ) fermion action on different size lattices. The temperature for each $\Nt$ was tuned by varying the lattice spacing $a$ through the gauge coupling $\beta$, and also for each lattice spacing the bare light and strange quark masses $m_l(a)$ and $m_s(a)$ were tuned so that the pseudo-Goldstone pion and kaon masses remain equal to the physical pion and kaon masses respectively. This scale illustrating the important Line of Constant physics\,\cite{Fodor2011lcp,Steinbrecher2018crossover} has been determined using the Sommer parameter $r_1$ as well as the kaon decay constant $f_K$, in which the quark masses are procured from the former scale while the latter is used for obtaining the temperature values already quoted in this paper. 
In this paper, we work on $32^3 \times 8$ lattices with physical values of bare quark masses in 2+1-flavor QCD, in which $m_l=m_s/27$. In order to calculate the $n$-point correlation functions $D_n$ and different order Taylor coefficients for each of our working temperatures, the first eight derivatives $D_1^f,\dots,D_8^f$ for each quark flavor $f=u,d,s$ were estimated stochastically using $500$ Gaussian random volume sources on every working gauge configuration, from which the $\muI$ derivatives were obtained. 

For this work, we have considered the isospin quantum numbers of $u\,,d$ quarks as $I_u=I_d=1/2$ resulting in the pion quantum number $I_\pi=1$. This subsequently gives the relation $\muI=(\mu_u-\mu_d)$ \,\,in our working baryon, strangeness, isospin (B,S,I) basis. A detailed derivation of this is presented in the Appendix \ref{appendix sec: fixing the convention}. This therefore leads to zero isospin correlation functions $D_n^I$ for all odd $n$, and real $D_n^I$ for even $n$ because of which, one finds no sign problem in lattice QCD simulations for finite real $\muI$. Importantly, this starting convention choice of the quantum numbers fixes the onset of pion condensation at $\muI=m_\pi$ as also shown explicitly in Ref.\cite{Andersen2023ofv}. This is unlike the same observed at $m_\pi/2$ in the work in \cite{Brandt2017isophdia}, where $I_u=I_d=1$ and $I_\pi=2$ have been used. Using this data and considering $20$K statistics for every temperature, we do our calculations of the cumulants and the Lee-Yang zeros of the isospin partition function, which constitute a central part of this paper. We demonstrate these results in the next section.

\begin{figure*}[ht]
    \begin{minipage}[t]{0.99\textwidth}
    \centering
    \includegraphics[width=.32\textwidth]{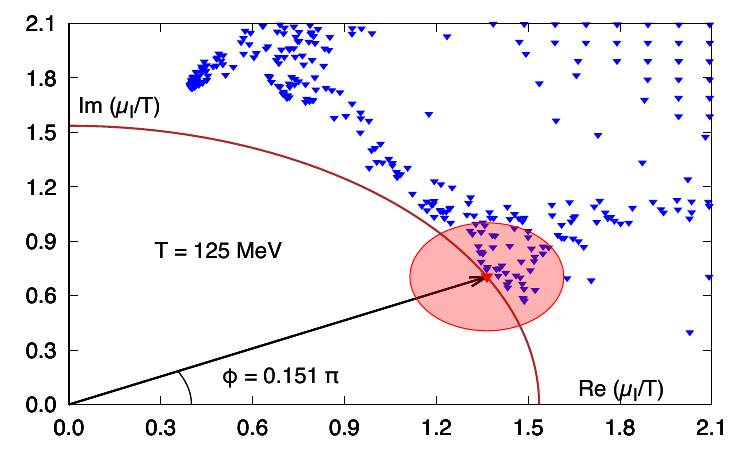}
    \includegraphics[width=.32\textwidth]{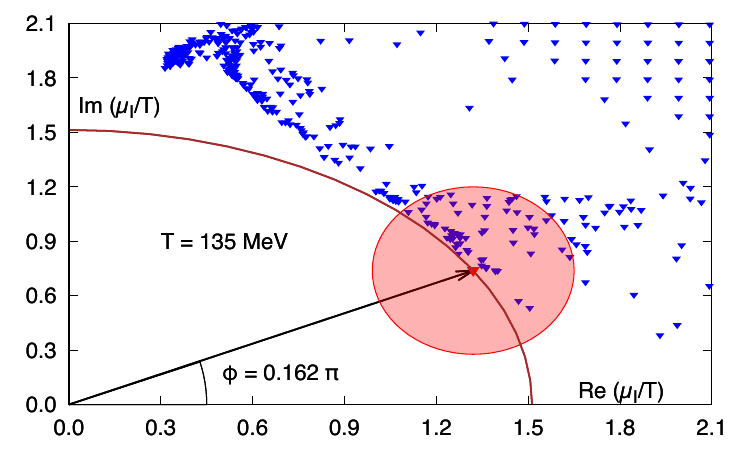} 
    \includegraphics[width=.32\textwidth]{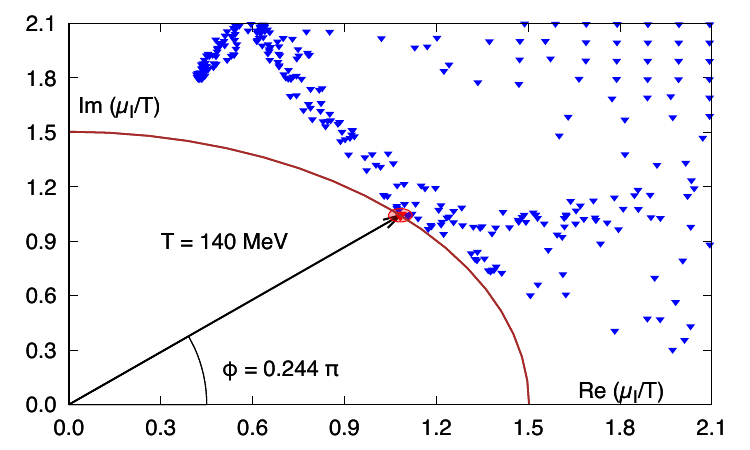} 
    \includegraphics[width=.32\textwidth]{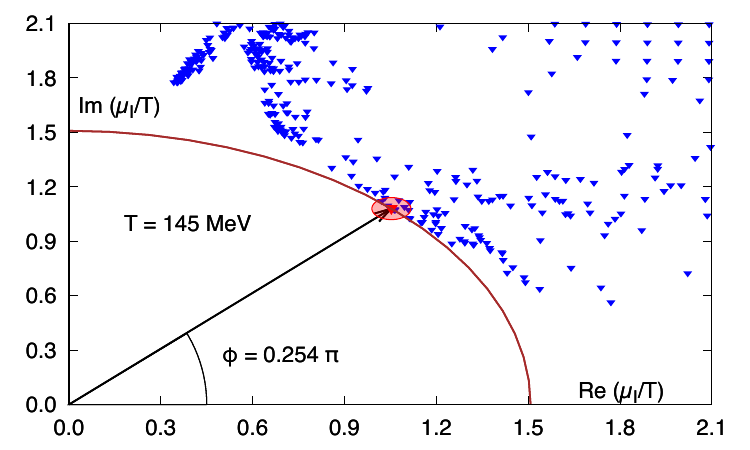} 
    \includegraphics[width=.32\textwidth]{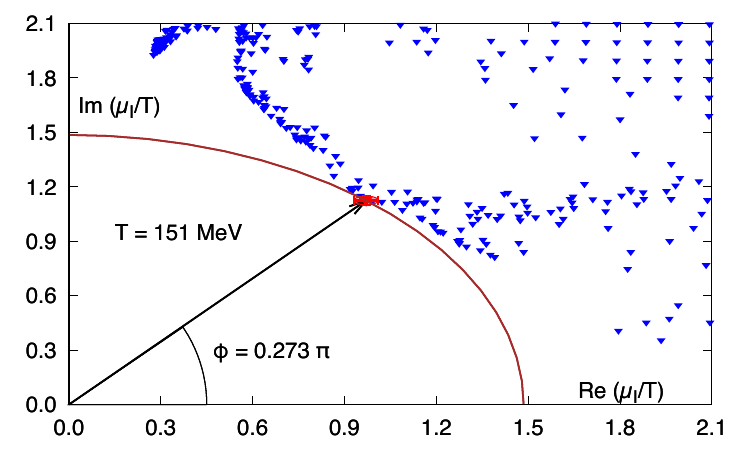}
    \includegraphics[width=.32\textwidth]{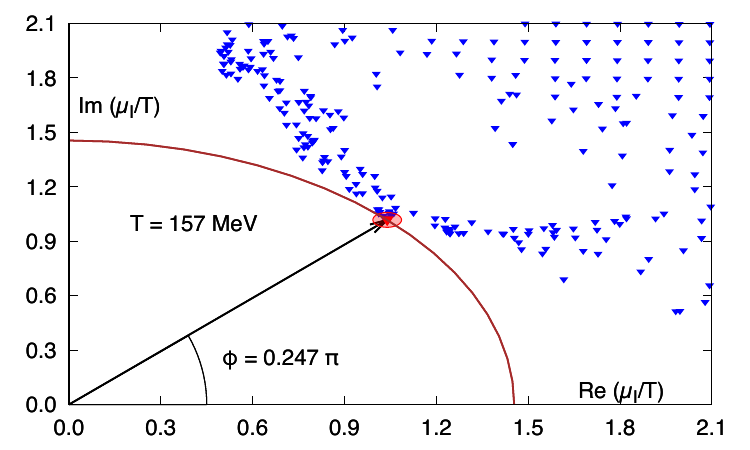} 
    \includegraphics[width=.32\textwidth]{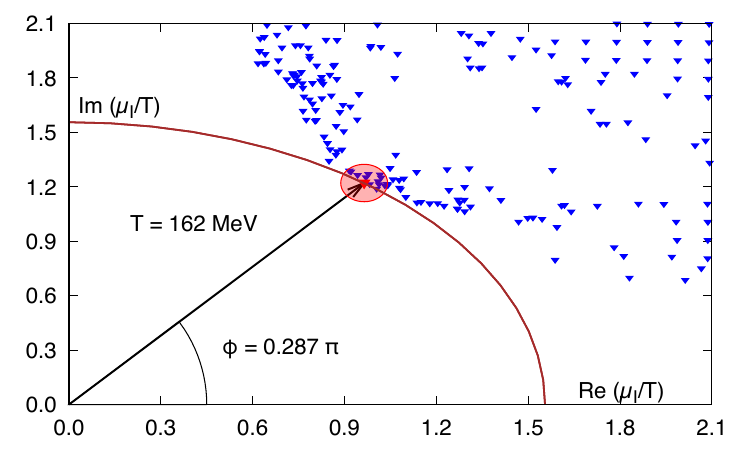}
    \includegraphics[width=.32\textwidth]{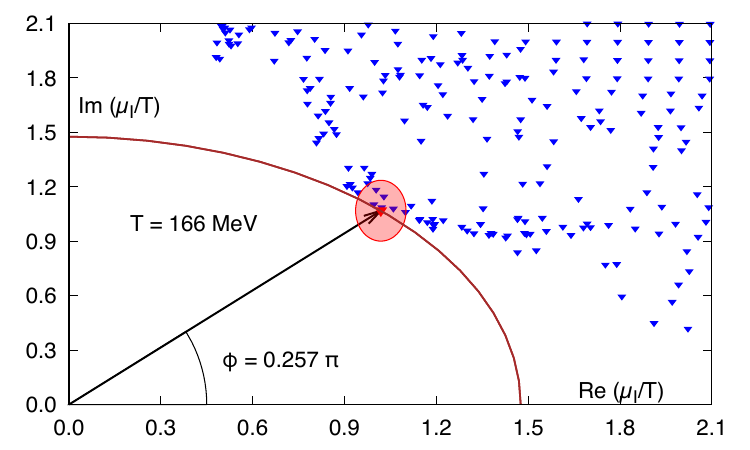} 
    \includegraphics[width=.32\textwidth]{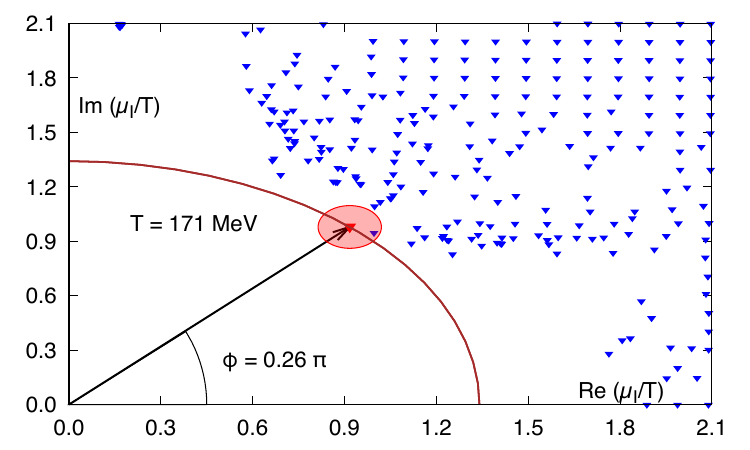} 
    \caption{Unbiased resummed estimates of the Lee-Yang zeros in complex $\muI$ plane (in blue) for $125 \leq T \leq 171$ (in MeV), making different angles $\phi$ at origin. The zeros closest to origin are shown for each $T$, with correlated errors along real and imaginary parts accounted by respective ellipses (in red). The brown lines show the circle of convergence in the first quadrant.}
    \label{fig:roots of Z}
    
   \end{minipage}
\end{figure*}

\section{Lee-Yang zeros of the isospin partition function}
\label{sec:muI}

\subsection{Observations and Results}
\label{subsec:LY zeros}

From this section onwards, we focus entirely on the isospin chemical potential $\muI$ discussion. 
We firstly compute the Lee-Yang zeros $\LeeYangzeroisospin$ of the QCD partition function $\partitionfunction$ in complex $\muI$ plane and present these results in Fig.\,\ref{fig:roots of Z}. 
As already mentioned, this $\partitionfunction$ is derived by approximating the exact path integral into a polynomial form using the unbiased exponential resummation formalism. 
 We compute these complex roots $\LeeYangzeroisospin$ of $\partitionfunction$ for different temperatures $T \in [125:171]$ MeV using the standard Newton-Raphson approach \cite{NewtonRaphson}, where the initial guesses of the roots are constructed considering $0 \leq \text{Re}(\muI)\,,\,\text{Im}(\muI) \leq 2.1$. This choice is based on the resulting observation, where we find the different-order Taylor and resummed series of the cumulants of $\partitionfunction$ start to deviate away from one another from $|\muI| \approx 1$, leading us to suspect that one may possibly encounter these zeros close to and beyond $|\muI| \approx 1$ in the complex $\muI$ plane. Also, this is related to our interest in finding the radius of convergence which is the distance of the nearest Lee-Yang zero from the origin. Also with this initial choice of complex $\muI$ range, we try ensuring that the guess values are not too different from the actual analytic roots, which is important to sustain the validity of the Newton-Raphson approximation used in this work. 
 Here in our work, the zeros are computed with error tolerance value $\epsilon=0.002$, with the upper bound $N_U$ on iterations kept sufficiently high $\sim \mathcal{O}(10^8)$. Obtained through repetitive trial and error methods, this large value of $N_U$ is to ensure that the algorithm does not fail and as mentioned before in Sec.\ref{subsec:Newton-Raphson}, performs all the necessary required iterations and evaluates $x_n$ and $x_{n+1}$, satisfying $\left|x_{n+1}-x_n\right| \leq \epsilon$. Surely, one can vary this tolerance value $\epsilon$ and observe if further smaller values ensure better results of $\muI^{LY}$ and offers better indications of the pion phase boundary and other aspects of the QCD isospin phase diagram. 
  However with this, one should also note the number of iterations increases and so does the subsequent time for convergence. The upper bound value $N_U$ therefore also requires attention, and possibly further increment to avert any possible lack of convergence of this algorithm. 
  
 These complex roots are obtained after being averaged over $N_B=100$ samples in each of which we bootstrap the original gauge ensemble using different random seed values, and thereupon evaluate these roots in the new bootstrapped gauge ensemble. With a total of $20$K configurations analysed in each such sample, this also allows one to calculate the respective errors in the real and imaginary part measurements of these zeros. In this work, we perform and illustrate this only for the respective closest zeros $\nearestLeeYangzeroisospin$ in Fig.\,\ref{fig:roots of Z} by constructing ellipses for different $T$, considering into account the correlations between the errorbars along the real $\quickLYRe=\nearestLeeYangzeroisospinreal$ and imaginary parts $\quickLYIm=\nearestLeeYangzeroisospinimag$ respectively. We also construct in this figure, the angle $\phi$  made at the origin by the complex zero $\nearestLeeYangzeroisospin$ for each working temperature. This is given by

\begin{equation}
    \phi=\tan^{-1} \left[\frac{\quickLYIm}{\quickLYRe}\right]
    \label{eq:phi}
\end{equation}
where $\phi \in [0:\pi/2]$ is ensured by the particle-antiparticle symmetry of QCD. 
Fig.\,\ref{fig:roots of Z} clearly illustrates that $\nearestLeeYangzeroisospin$ approaches the real $\muI$ axis as $T$ is lowered, specially in the QCD hadronic regime for $T<T_{cr} \sim157$ MeV. Except for $162$ MeV, we notice a similar sort of behaviour of $\nearestLeeYangzeroisospin$ for the high $T$ plasma regime too. These are well-indicated by the reducing $\phi$ values with decreasing temperature in both the QCD phases on either sides of crossover temperature $T_{cr}$. 
 We also apparently notice that with increasing $T$ values, the rectangular grid of blue points initially starting from the top right corner of the plots continues to become enhanced and enlarged, resulting to larger density of zeros in the plots. This is obvious, as the Newton-Raphson method fails to converge due to very large values of the exponentially resummed $\partitionfunction$. This manifests in the regime of large complex $\muI$ values as well as for higher values of $T$, where the $n$-point correlation functions increase, resulting to loss of convergence within a few iterations and hence, the obtained estimate value is not far away from the initial guess value $\mu_0$ in the complex $\muI$ plane. 

 \begin{figure*}[ht]
    \begin{minipage}[t]{0.99\textwidth}
    \centering
    \includegraphics[width=0.32\textwidth]{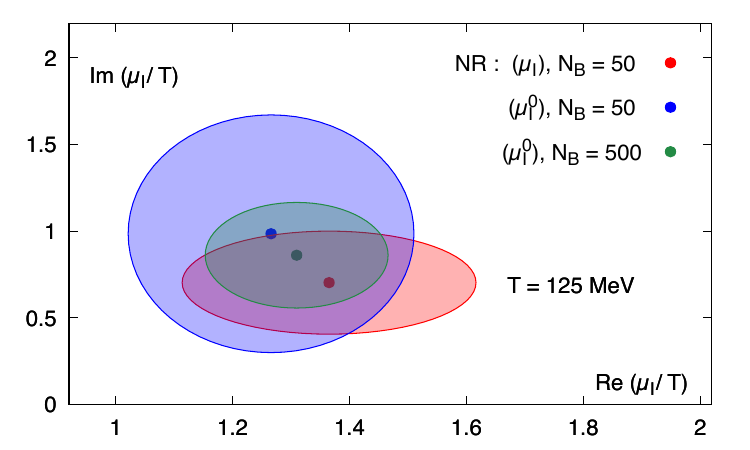}
    \includegraphics[width=0.32\textwidth]{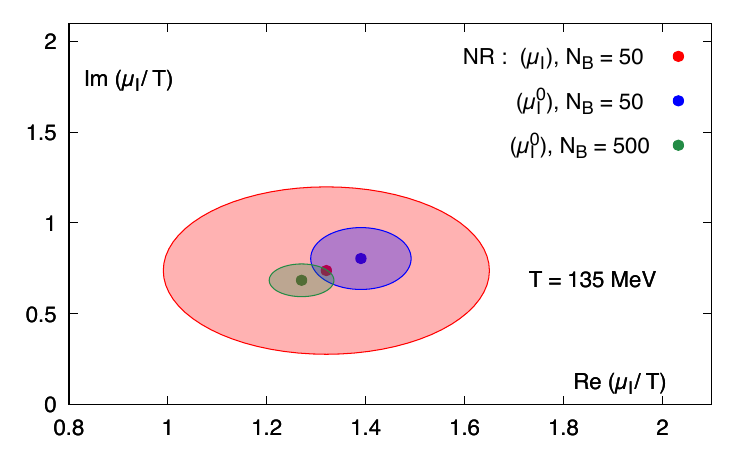}
    \includegraphics[width=0.32\textwidth]{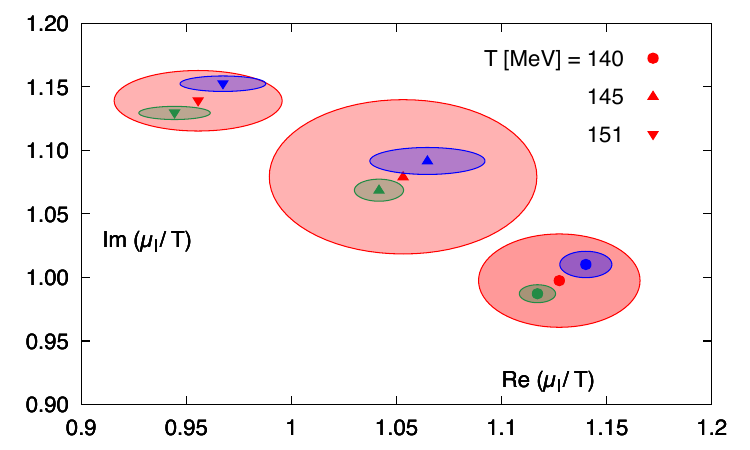}
    \caption{Estimates of the Lee-Yang zeros in complex $\muI$ plane computed using the Newton-Raphson method are shown in red, blue and orange points respectively for $T=125$ MeV (left) and $135$ MeV (center). See full text for details. (Right) The same plot now for all the other three temperatures, $140 \leq T \leq 151$ MeV in the hadronic regime.}
    \label{fig:stability}
     \end{minipage}
\end{figure*}

 Interestingly we also find from Fig.\,\ref{fig:roots of Z}, that the resummed radius of convergence $\ROCresum = |\nearestLeeYangzeroisospin|$ remains close to around $1.5$ within errorbars, which we discuss in Sec.\ref{sec:Radius of convergence}.  Defined as $\ROCresum = \ROCTresum/T$, this suggests that $\muI^{\rho}$ should show a monotonic behaviour in $T$, possibly in a linear manner in this regime. With $\muI^\rho$ coinciding with the pion phase boundary in the isospin phase diagram shown before in \cite{Borsanyi2023tdp}, this observation goes well otherwise with the predictions of $\chi$PT \cite{Son2000xc}, stating a similar monotonic behaviour of $\mu_{I,c} \sim \muI^\rho$ with $T$ in this regime of the present work, where $\muI \ll m_\rho$\footnote{rho meson mass $m_\rho=770$ MeV} ensuring the validity of $\chi$PT. However before believing these results and drawing some possible conclusions, it is important to check the stability of the algorithm and the reliability of these calculated Lee-Yang zeros, which we discuss in the following section.

\subsection{Stability of the results}
\label{subsec:stability}

In this section, we comprehensively check the stability of the closest Lee-Yang zero estimates, which we have illustrated in the previous section and are also shown in Fig.\ref{fig:roots of Z}. For this, we consider these estimated zeros in Fig.\ref{fig:roots of Z} at different $T$ as our initial guesses and subsequently implement the Newton-Raphson algorithm on these respective zeros for different temperatures. The consequent results  are outlined in Fig.\ref{fig:stability}. With these new starting points $\mu_0$ along with same values of tolerance $\epsilon$ and maximum number $N_{U}$ of allowed iterations as before, we use Newton-Raphson method and re-calculate using identical number of bootstrap samples $N_B$ in order to check if the algorithm manages to converge to and identify these zeros efficiently. In Fig.\ref{fig:stability}, we clearly observe that for all the five working temperatures, $125 \leq T \leq 151$ MeV, the final estimates of the zeros agree considerably well with the starting zero estimates within respective errorbars. We also show a zoomed version of these results for the lowest two working temperatures at $T=125\,,135$ MeV in the same Fig.\ref{fig:stability}. These figures very apparently illustrate that this agreement holds true even when the bootstrap samples $N_B$ are increased by an order from $50$ to $500$, even though one can also observe that the errors reduce for the latter with higher $N_B=500$, as outlined by the green elliptical regions in Fig. \ref{fig:stability}.

\begin{figure}
    \centering
    \includegraphics[scale=0.6]{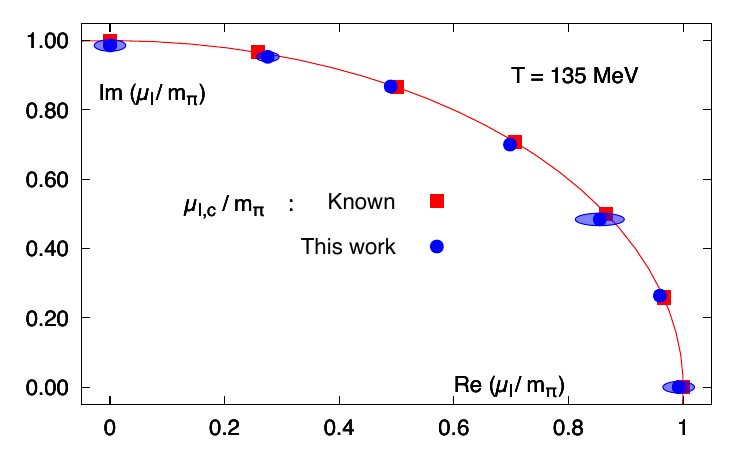}
    \caption{Estimates of Lee-Yang zeros at $T=135$ MeV, when one considers as initial guesses, the values of complex $\muI$ satisfying $\left|\muI\right|=m_\pi$. The red and blue points respectively showcase the initial guess values and the final Newton-Raphson estimates obtained with these initial guess values. Refer to the full text for details.}
    \label{fig:Bastian-135}
\end{figure}

In addition to demonstrating this, we also further perform an extension of this study, by making use of the known critical points at $135$ MeV and thereupon implementing our working Newton-Raphson method on them for the purpose of crosschecking the reliability of our calculated Lee-Yang zeros, shown already in Fig.\ref{fig:roots of Z}. The results of this stability check are presented in Fig.\ref{fig:Bastian-135}.
The complex critical points $\mu_{I,c}$ as the corresponding starting guesses used in this section, have been constructed parametrically using the following : 

\begin{equation}
\text{Re}\,(\mu_{I,c})=m_\pi\,\cos{\phi}\hspace{.3cm},\hspace{.3cm} \text{Im}\,(\mu_{I,c})=m_\pi\,\sin{\phi}
\label{eq:parameteric}
\end{equation}
 where $\phi$ is given in Eqn.\eqref{eq:phi}, and $\muI=m_\pi$ is the onset point of the pion condensation based on the starting convention of this work, also used and outlined before in \cite{Andersen2023ofv}. The choice of $T=135$ MeV is motivated by \cite{Brandt2017isophdia}, where the pion condensation line is predicted at $\muI=m_\pi/2$ for $T \leq 140$ MeV. It must be noted that this factor of $1/2$ in \cite{Brandt2017isophdia} arises purely due to a different starting convention and therefore, differs also from other related works \cite{Son2000xc,Andersen2023ofv}. These works not only obtain the onset at $\muI=m_\pi$ using chiral perturbation theory, but also use the exactly same convention as used in this present work. A relevant discussion about this is given in Appendix \ref{appendix sec: fixing the convention}. 
 
 In this section, we consider seven critical points satisfying Eqn.\eqref{eq:parameteric} outlined as red points in Fig.\ref{fig:Bastian-135}. These correspond to seven values of $\phi$, namely $\phi=n\,\pi/12$ with integer $n$ satisfying $0 \leq n \leq 6$. In Fig.\ref{fig:Bastian-135} (blue points and ellipses), we observe that the final zero estimates obtained as a result of using the Newton-Raphson method on these red points as initial guesses, agree appreciably well within errors, and converge commendably with these red points. Although interesting future works would involve improvement in the accuracy and precision of this agreement as well as associated manifestations for smaller tolerance values, we believe that the present level of agreement achieved is appreciable and reliable enough for the current paradigm of this work.

\section{Real and imaginary parts of Lee-Yang zeros}
\label{sec:Real and imaginary}

\begin{figure}[ht]
    \centering
    \includegraphics[scale=0.6]{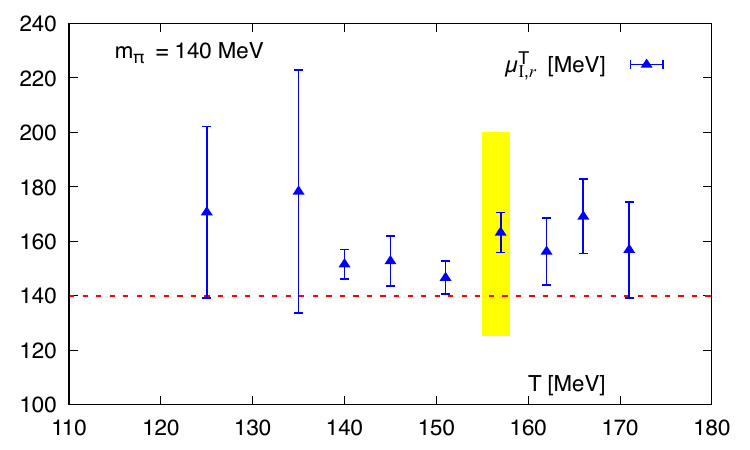}
    \includegraphics[scale=0.6]{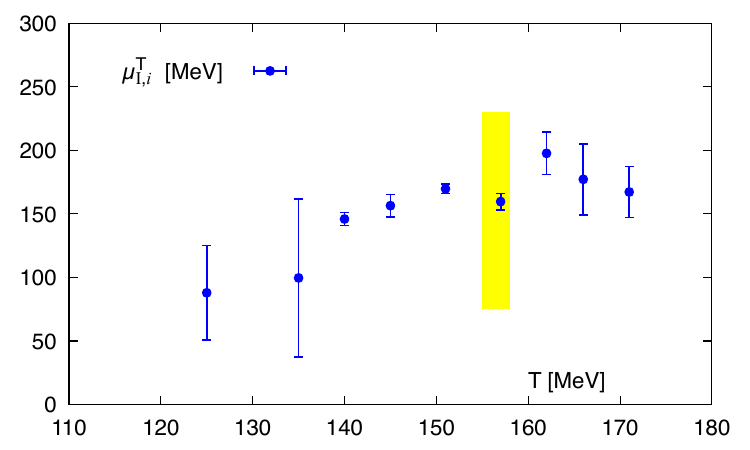}
    \caption{(Top) The real part $\quickLYReT=\nearestLeeYangzeroisospinreal \cdot T$ \,\,and (Bottom) the imaginary part $\quickLYImT=\nearestLeeYangzeroisospinimag \cdot T$  as function of temperature $T$ for $125 \leq T \leq 171$ MeV with errorbars. The red dotted line in the top figure marks the value of physical $m_\pi=140$ MeV. The yellow band provides the crossover region $T_{cr}=156.5\,(1.5)$ MeV, as obtained from \cite{HotQCD2014QCD}.}
    \label{fig:Real and imaginary parts of zeros}
\end{figure}

We explore the real and imaginary parts of these zeros separately and probe their behaviour as a function of the temperature. Even though we also evaluate these zeros for higher temperatures in the so-called the quark gluon plasma regime above $T_{cr} \sim 157$ MeV, we mostly concentrate on the lower temperatures belonging in the hadronic regime, primarily because of our interest in knowing the behaviour of these zeros for lower temperatures towards $T=0$ and eventually aiming to map a critical point on the pion phase critical line in the low $T$, small $\muI$ regime of the QCD isospin phase diagram. 

Having verified the reliability of the current mathematical setup for this work, we turn our attention in this section towards the temperature-dependent behaviour of the real and imaginary parts of nearest Lee-Yang zeros $\nearestLeeYangzeroisospin$ namely, $\quickLYRe$ and $\quickLYIm$ in the complex $\muI$ plane. These are demonstrated in Fig.\ref{fig:Real and imaginary parts of zeros}. One clearly observes a finite value of $\quickLYImT$ in this figure even at lowest working $T=125$ MeV, where $\quickLYImT$ is given as $\quickLYImT=\quickLYIm \cdot T$. This is not surprising, as it is a natural expectation in a finite volume. However strictly speaking, one needs to approach the thermodynamic limit in this regard which is to check if this value of $\quickLYImT$ remains finite or it vanishes in this limit. While the former would indicate a genuine crossover, the latter would imply a true phase transition at this temperature and hence this is definitely an interesting prospect to probe into in the future. We clearly find in Fig.\ref{fig:Real and imaginary parts of zeros}, that $\quickLYIm$ reduces monotonically with decreasing temperature $T$ at least in the low $T$-hadronic regime below $T_{cr} \approx 157$ MeV. Aside this, one also observes that the associated errorbars of these $\quickLYImT$ values increase with similar statistics for $T\leq 135$ MeV, which can possibly indicate existence of the pion condensate phase and related enhancement of fluctuations among the $\quickLYReT\,, \quickLYImT$ values across different bootstrap samples. One definitely needs to perform further crosschecks and also explore other associated relevant observables for a more firm and conclusive statement in this regard. 
We also observe in Fig.\ref{fig:Real and imaginary parts of zeros}, that the leading Lee-Yang zero exhibits a small yet finite imaginary part across the explored temperature range. While this observation specially for $T \leq 140$ MeV in this figure may seem at odds with the usual, prevailing expectation of a real zero for a genuine second-order phase transition in the thermodynamic limit as provided by the current state-of-the-art studies\,\cite{Brandt2017isophdia}, we interpret this feature as a combined outcome of finite-volume effects and possible artifacts of the working method of unbiased exponential resummation in this present work. With the belief that a comprehensive finite-size scaling analysis of these zeros and their behaviour in thermodynamic limit would explain these non-vanishing $\quickLYIm$ and possibly mitigate them for $T \leq 140$ MeV, we leave a detailed discussion of these various artifacts and corrections for future.

Apart from the behaviour of the imaginary part $\quickLYImT$, Fig.\ref{fig:Real and imaginary parts of zeros} also demonstrate the $T$-dependence of the real part $\quickLYRe$ or equivalently $\quickLYReT=\quickLYRe \cdot T$. Unlike $\quickLYImT$,  one observes in this case that, $\quickLYReT$ qualitatively exhibits a fluctuating yet saturation-like behaviour, with no indications of any sort of strict monotonic reduction or increment like $\quickLYImT$. Interestingly we find in this case, that at least for lower $T \leq 135$ MeV i.e. at $T=125\,, 135$ MeV, $\quickLYReT$ shows good agreement within errorbars with $\mu_{I,r}=m_\pi=140$ MeV which also marks the value of the physical pion mass used in this work. This is depicted in Fig.\ref{fig:Real and imaginary parts of zeros} outlined by the red dotted line. Besides these two temperatures, the behaviour of $\quickLYReT$ for other temperatures, also in the hadronic regime for $140 \leq T \leq 151$ MeV in Fig.\ref{fig:Real and imaginary parts of zeros} clearly manifests values, which remain quantitatively close to $140$ MeV and therefore do not differ too much from the red dotted line. We reserve the discussions for the high $T$ regime for the future. 

These joint behaviour of $\quickLYReT$ and $\quickLYImT$ for the lower $T \leq 135$ MeV, go well and shows consistency with the present idea outlined already in \cite{Borsanyi2023tdp}, where it is argued that the radius of convergence $\muI^\rho$ for a finite volume at large enough orders, satisfies

\begin{figure}[ht]
    \centering
    \includegraphics[scale=0.6]{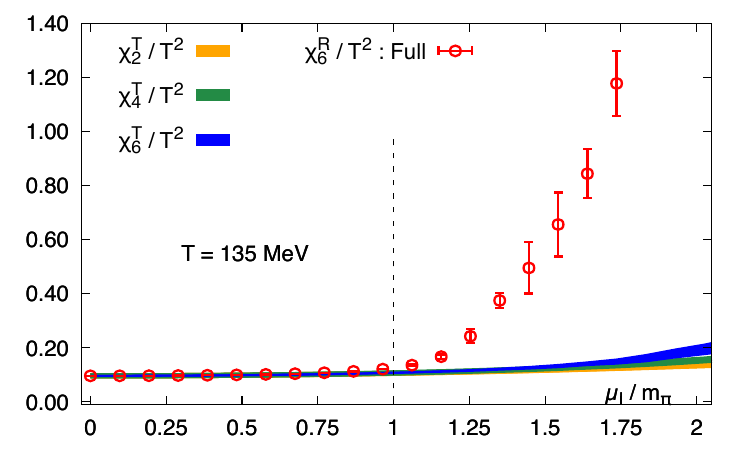}
    \caption{Figure showing the behavior of the number-density to isospin chemical potential ratio {\it i.e.} isospin susceptibility $\chi_I$ as a function of $\muI$ for different order Taylor series (in bands) and for the full reweighting by unbiased exponential resummation (as red circular points). The dotted vertical line resembles the $\muI=m_\pi$ line.}
    \label{fig:increasing number density ratio}
\end{figure}

\vspace{-.4cm} 

\begin{equation}
\muI^\rho \approx \sqrt{\left(\mu_{I,p}\right)^2 + \left(\quickLYImT\right)^2} \,,   
\label{eq:ROC in Borsanyi}
\end{equation} 
with $\quickLYImT$ the imaginary part  of the closest complex Lee-Yang zero and $\mu_{I,p}$ the phase transition chemical potential value in the infinite-volume thermodynamic limit. 
This implies that the real part $\quickLYReT$ approximately saturates towards the expected true phase transition $\mu_{I,p}$ even at finite volume for low enough temperatures, where one expects the imaginary part $\quickLYImT$ non-vanishing, finite and showing some temperature-dependent behaviour. We observe both of these manifestations from Fig.\ref{fig:Real and imaginary parts of zeros} within errorbars, where there is a monotonic reduction with reducing $T$ in the imaginary sector and the real parts seem to saturate towards the physical pion mass value $140$ MeV, shown by the red dotted line. Consequently, the failure of the radius of convergence estimate $\muI^\rho$ in Eqn.\eqref{eq:ROC in Borsanyi} to converge at the true critical point in finite-volume simulations for these temperatures is mostly attributed to non-vanishing value of the imaginary part $\quickLYImT$ of the Lee-Yang zero $\muI^0$, which may possibly vanish in thermodynamic limit if the phase transition is genuine for these range of temperatures. We leave this analysis for a possible future work.

In line with \cite{Borsanyi2023tdp} and also as some evidence of the onset of pion condensate phase at $\muI=m_\pi$, we present Fig.\ref{fig:increasing number density ratio}, where we outline the isospin susceptibility $\chi_n$ behaviour for different orders $n$ at $T=135$ MeV as a function of $\muI/m_\pi$ with $m_\pi=140$ MeV. This is to obtain possible set of observations from the perspective of reweighting via unbiased exponential resummation approach. The isospin susceptibility $\chi =\partial \mathcal{N}/\partial \muI$ is chosen, as it is roughly the number-density to chemical potential ratio probed in \cite{Borsanyi2023tdp}. Fig.\ref{fig:increasing number density ratio} vividly illustrates that starting from $\muI=m_\pi$, the sixth order reweighted isospin susceptibility $\chi_6^R$ obtained via unbiased exponential resummation method shown as red points, deviates sharply away from the same obtained from different order Taylor series $\chi_n^T$ shown as bands in this figure. This is encouraging, as it offers similar manifestations to what has also been observed in \cite{Borsanyi2023tdp}, and thus provide promising indications about possible existence of the pion condensation phase boundary at $\muI=m_\pi$ for this temperature.

\section{Radius of convergence}
\label{sec:Radius of convergence}

 Having outlined the behaviour of real and imaginary parts of closest Lee-Yang zero $\nearestLeeYangzeroisospin$ as functions of $T$, we outline the radius of convergence estimate provided by $\nearestLeeYangzeroisospin$ for various working $T$ in this section. Besides presenting some comparative studies, we also attempt to obtain a possible critical point estimate on the pion condensate phase boundary in the isospin phase diagram via linear extrapolation which we detail in this section.

\subsection{Results : From Linear extrapolation Fits}
\label{subsec:estimates from extrapolation of zeros for lower T}

\begin{figure}[ht]
    \centering
    \includegraphics[scale=0.6]{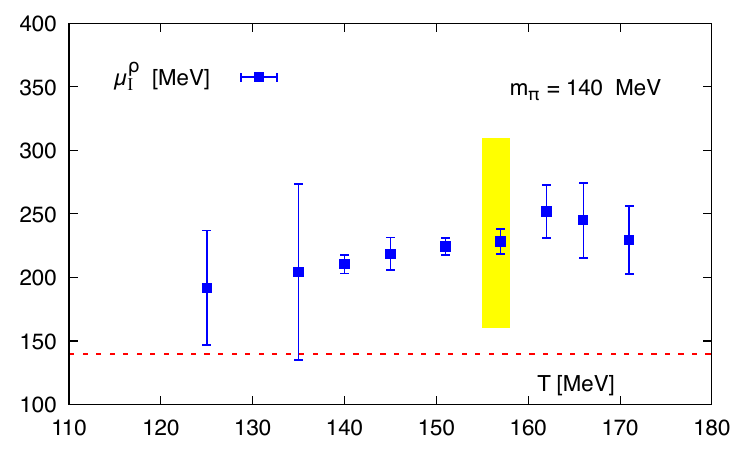}
    \caption{The radius of convergence $\muI^{\rho}$ in units of MeV (in blue points) as function of temperature $T$, for $m_\pi=140$ MeV}
    \label{fig:ROC}
\end{figure}

In this subsection, we plan to outline two things as follows : (1) the radius of convergence $\muI^{\rho}$ as a function of temperature $T$ and (2) a possible estimate of a critical point on the second order critical pion phase boundary, which we aim to attain by means of a linear extrapolation of the $\quickLYImT$ data points as a function of $T$, shown already before in Fig.\ref{fig:Real and imaginary parts of zeros}. As known from previous works, the radius of convergence of the Taylor expansion of free energy or equivalently, the free energy per unit volume \textit{i.e.} pressure can offer a reliable estimate of the critical point. This coincides with the distance of the nearest Lee-Yang zero of the partition function from the origin in the complex $\mu$ plane ($\mu=\muI$ here). As mentioned before, Fig.\ref{fig:ROC} illustrates the $T$-\,dependence of the physical radius of convergence $\ROCTresum$ in energy (MeV) units, which very understandably manifests a monotonic reduction of $\muI^\rho$ with lowering $T$ in the hadronic regime below $T_{cr} \sim 157$ MeV. Besides this, one also observes in this figure that for low $T \leq 135$ MeV, these estimates come close and agree within errorbars with $\muI^\rho=140$ MeV shown by the red dotted line  which as mentioned before, is the pion mass value used in this work. Unlike Ref.\cite{Brandt2017isophdia} which outlines a faster saturation of $\muI^\rho$ towards $\muI=140$ MeV except the choice of a different starting convention, the present observation of Fig.\ref{fig:ROC} yields a more gradual and monotonic nature in the decreasing values of $\muI^\rho$ with reducing values of $T$, which aligns better and consistent with the predictions of the $\chi$PT in \cite{Son2000xc}. These warrant further detailed works for obtaining  enhanced definite conclusions. 

\begin{figure}[ht]
    \centering
    \includegraphics[scale=0.6]{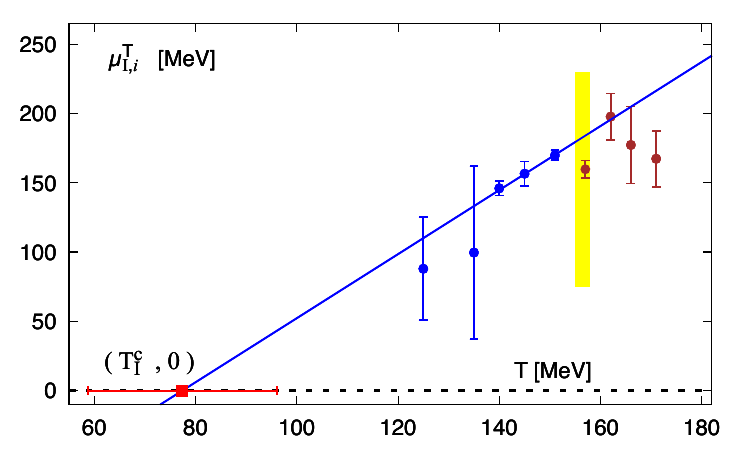} \\
    \includegraphics[scale=0.6]{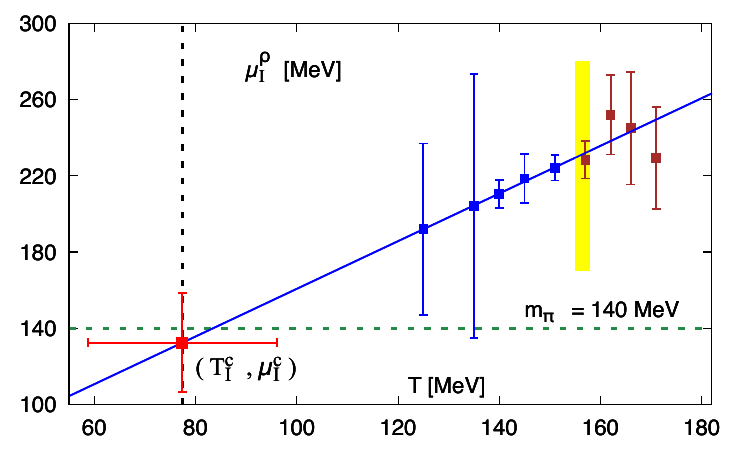}
    \caption{Plots showing linear fitting and extrapolation to the data points concerning (Top) $\quickLYImT$ and (bottom) $\muI^\rho$. The horizontal and vertical errorbars (in red) illustrate the errors in $\crittemp$ and $\critmu$ respectively. See text for other details.}
    \label{fig:fitting}
\end{figure}
 
For the critical point determination, we follow the usual state-of-the-art approach as followed before in numerous works involving Lee-Yang zeros \cite{Dimopoulos2021vrk,Schmidt2022ogw,Clarke2024seq}. The central idea is to evaluate the temperature $\crittemp$ at which the nearest complex Lee-Yang zero becomes purely real and its imaginary part vanishes. With this value of $\crittemp$, we then procure the value of $\critmu$ which is $\muI^\rho (T)$ at $T=\crittemp$ and subsequently the physical critical point  is given by $\left(\crittemp, \critmu\right)$ in the $T-\muI$ plane. 
With no working data available at present for further lower $T<125$ MeV to outline the behaviour of $\quickLYImT$ and $\ROCTresum$ for lower $T$ where one expects the former to vanish for some finite value of $T=\crittemp$, we resort to performing a linear extrapolation to these data points shown before in Figs.\ref{fig:Real and imaginary parts of zeros} and \ref{fig:ROC} respectively. This is also based on the observations from these figures, manifesting monotonic behaviour of $\quickLYImT$ and $\muI^\rho$ vs $T$ in the hadronic regime. Despite having no notion of crossover in this probing regime of $T$ and $\muI$ as per the current state-of-the-art isospin phase diagram, this extrapolation shown in Fig.\ref{fig:fitting} is carried out entirely with the purpose of determining an estimate of the critical temperature $T_I^c$ within the present setup and limitations of this work. Without intending to locate the $T_I^c$ in the thermodynamic limit which we leave for a promising work in the future involving detailed volume dependence studies of these zeros, this extrapolation rather attempts here to illustrate the temperature-dependent trend in the behavior of these leading zeros and by following this trend, determine $T_I^c$ at which the Lee-Yang zero becomes real with vanishing $\quickLYIm$.  

We demonstrate the fit results explicitly in Fig.\ref{fig:fitting} where we consider as input data to the fitting, only the blue data points lying in the low $T$ hadronic regime below $T_{cr} \sim 157$ MeV shown by the yellow band. The data points in the plasma regime for $T > T_{cr}$ outlined in brown, are excluded and not considered for the linear fitting and extrapolation. Using linear ansatz of the forms $\quickLYImT = a\,T+b$ and $\muI^\rho=c\,T+d$ with fitting parameters $a,b,c,d$, we employ error propagation and least $\chi^2$ fitting method and obtain the following estimates in MeV units: 


\begin{figure*}[ht]
    \begin{minipage}[t]{.99\textwidth}
    \centering
1    \includegraphics[width=.32\textwidth]{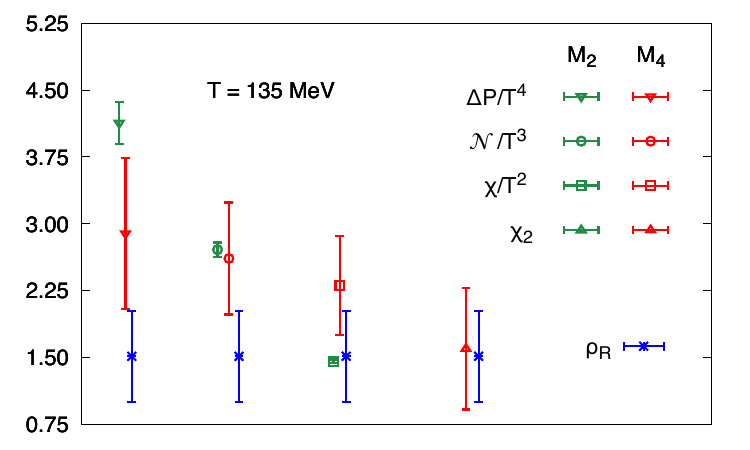}
    \includegraphics[width=.32\textwidth]{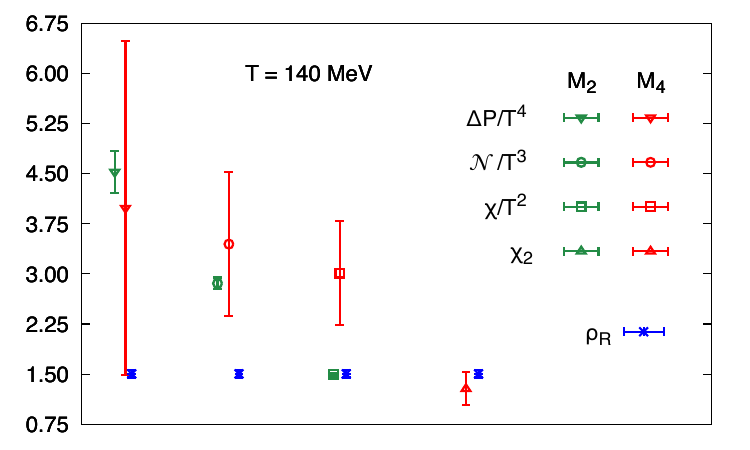}
    \includegraphics[width=.32\textwidth]{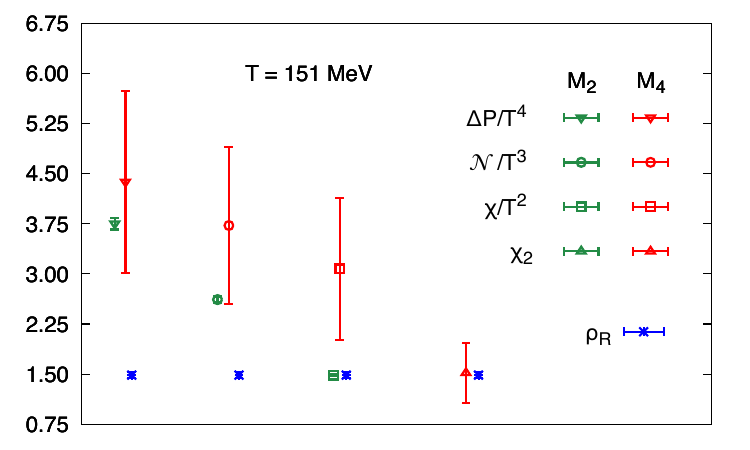}
    \includegraphics[width=.32\textwidth]{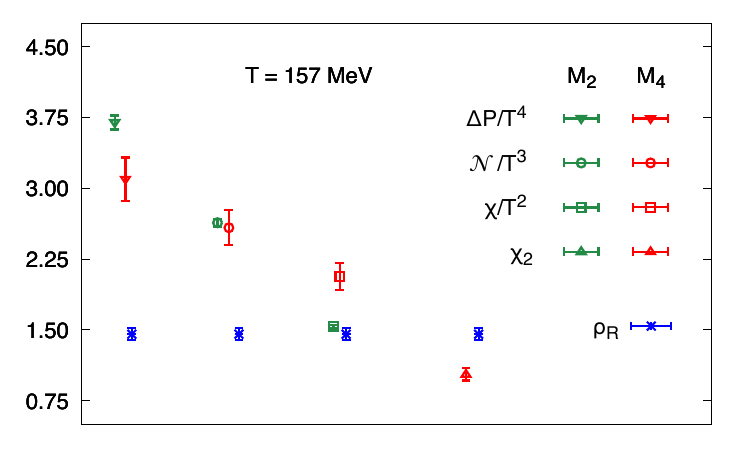} 
    \includegraphics[width=.32\textwidth]{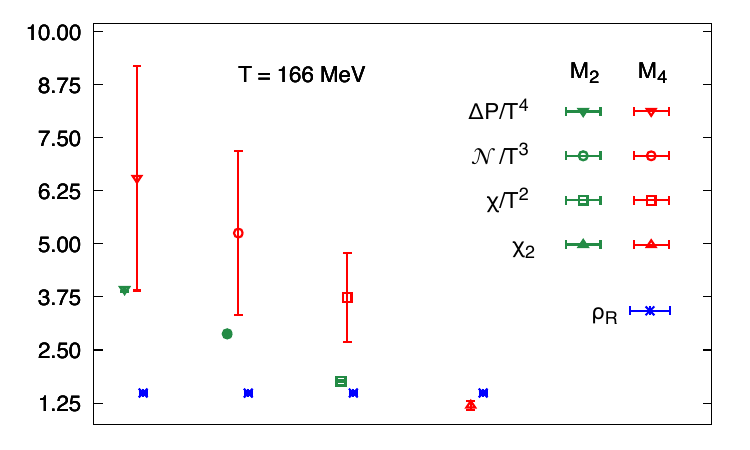}
    \includegraphics[width=.32\textwidth]{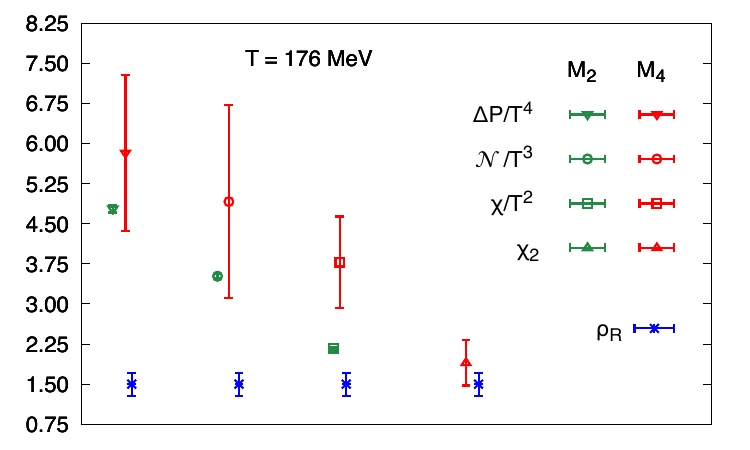}
    \\
    \caption{The Mercer-Roberts estimates $\M_n$ for $n=2,4$ obtained from the Taylor series of the first four cumulants $\Delta P$, $\mathcal{N}$, $\chi$, $\chi_2$ for different $135 \leq T \leq 176$ MeV, shown by the green and red points. The blue points indicate resummed estimate $\ROCresum$.}
    \label{fig:MR}
    \end{minipage}
\end{figure*}

\begin{align}
    &\crittemp=77.38 \pm 18.71 \notag \\
    &\critmu=132.40 \pm 25.94 
    \label{eq:critical point}
\end{align}
From the above calculation, we find small values of $\chi^2/dof$ around $0.7$ and $0.02$ for $\crittemp$ and $\critmu$ respectively, which we think are possibly due to the large errorbars for $T=125$ and $135$ MeV in Figs.\ref{fig:Real and imaginary parts of zeros} and \ref{fig:ROC}. The larger errors for these temperatures contribute also to sizable errors in respective $\crittemp$ and $\critmu$ estimates, and thus require more attention. In this regard, one can think of future works involving higher number of statistics as well as bootstrap samples $N_B$. One can also attempt to work with further higher order of unbiased resummation which can possibly allow us to control these errors in a better fashion by eliminating higher order biased contributions in $\muI$.

 Nevertheless, this present estimate $\left(\crittemp,\critmu\right)$ in above Eqn.\eqref{eq:critical point} of a possible critical point obtained for the first time from unbiased exponential resummation method, agree commendably well within the respective errorbars with the predicted second order pion condensation phase boundary, which at this lower $T \approx 80$ MeV should correspond to $\muI=m_\pi$ in the isospin phase diagram in the $T$-$\muI$ plane where $m_\pi$ is equal to approximately $140$ MeV, given this work is performed considering physical values of up and down $(u,d)$ light quark masses in ($2$+$1$)-flavor QCD. This current stature of the pion condensate critical line, including its existence for low $T,\muI$ and its onset from $\muI=m_\pi$ at least for $T \leq 140$ MeV with the present working convention is unanimously supported by present state-of-the-art related works \cite{Andersen2023ofv,Brandt2017isophdia,Brandt2019isophdia}. A positive feature of this estimate is that it is obtained from a direct Lee-Yang zero analysis of the isospin partition function without the requirement of calculating cumulants and their subsequent Lee-Yang singularities which have been done in some previous works \cite{Schmidt2024uuk,Schmidt2025ftp}. Another positive achievement is that, this result is attained naturally as an outcome of the simple linear extrapolation in this work without the need of any other ansatz that relied on apriori assumption about the properties of this pion condensation phase like its order, universality class and point of onset in the isospin phase diagram as possible input to the associated relevant fitting parameters, and the fitting itself. This is significant and promising, despite the fact that this work in its present form does not infer anything about the order and universality class of this phase transition. 
Note, the visible discrepancy between our obtained $T_I^c$ estimate in Eqn.\eqref{eq:critical point} and the same in \cite{Brandt2017isophdia} is because of different possible corrections contributed by the finite-volume effects of our working lattice as well as the artifacts of the working method, which also cause finite $\quickLYIm$ values for $T \leq 140$ MeV as seen in Fig.\ref{fig:Real and imaginary parts of zeros} and clarified in Sec.\ref{sec:Real and imaginary}. To resolve these and understand their severity, one clearly requires simulations at higher lattice volumes and explore volume dependence of respective $\quickLYRe, \quickLYIm$ of these zeros, by approaching thermodynamic limit. As per the existing literature \cite{Brandt2017isophdia,Borsanyi2023tdp}, one expects observing real zeros with vanishing $\quickLYIm$ for all $T \leq 140$ MeV and this is something we look to investigate thoroughly in a future work as previously mentioned. In spite of these existing limitations to be explored further in future, the positive upshot of this work is the successful mapping of a genuine critical point  (Eqn.\eqref{eq:critical point}) on the second order pion condensate critical line in the isospin phase diagram, using this Lee-Yang zero analysis. 
  
  Having this resummed estimate $\muI^\rho$ of radius of convergence, we are now in a position to illustrate some comparisons which we do in the following sections.

\subsection{Comparisons with Mercer-Roberts (MR) estimates}
\label{subsec: Ratio and MR estimates}

In this section, we compare this resummed estimate of the dimensionless radius of convergence $\ROCresum=\muI^\rho/T$ with the Mercer-Roberts (MR) estimates of the Taylor series. In relation to Taylor series, this is important for determining the extent of reliability of this series approximation of the cumulant measurements in $\muI$. Given the Taylor coefficients $c_n=\susc_n/n!$, the $n^{\text{th}}$ order estimator for the MR method of estimation is given by : 
\begin{equation}
    M_n = \left|\frac{c_{2n+2}\,c_{2n-2} - c_{2n}^2}{c_{2n+4}\,c_{2n} - c_{2n+2}^2}\right|^{1/2},
    \label{eq:MR}
\end{equation}
where the different-order estimates converge to the true radius of convergence $\rho$ in $n \to \infty$ limit. We outline this comparison in Fig.\,\ref{fig:MR} respectively, where we show the resummed radius of convergence $\rho_R$ and the Mercer-Roberts estimates $M_n$ for different orders $n$ using results of the Taylor coefficients $c_n$ derived from the first three cumulants of the partition function, which are excess pressure $\Delta P$, number density $\mathcal{N}$ and the susceptibility $\chi$. An important aspect to note in this regard, is the familiar ratio estimates $\rho_n$\,\footnote{$\rho_n=\sqrt{c_{2n}/c_{2n+2}}$} of different orders $n$ are not considered here for comparison. This is because, they do not converge for $n\to \infty$ limit owing to the non-vanishing, finite imaginary parts of the leading complex Lee-Yang zeros for these temperatures. This feature has been detailed mathematically in \cite{Pasztor2019mercer}. Also previous studies \cite{Mercer1990,Vovchenko2017cluster} have showed the Mercer-Roberts estimates exude better order-by-order stability, with more reliable convergent properties over the conventional ratio estimates $\rho_n$.
\begin{figure*}[ht]
    \begin{minipage}[t]{.99\textwidth}
    \centering
    \includegraphics[width=.32\textwidth]{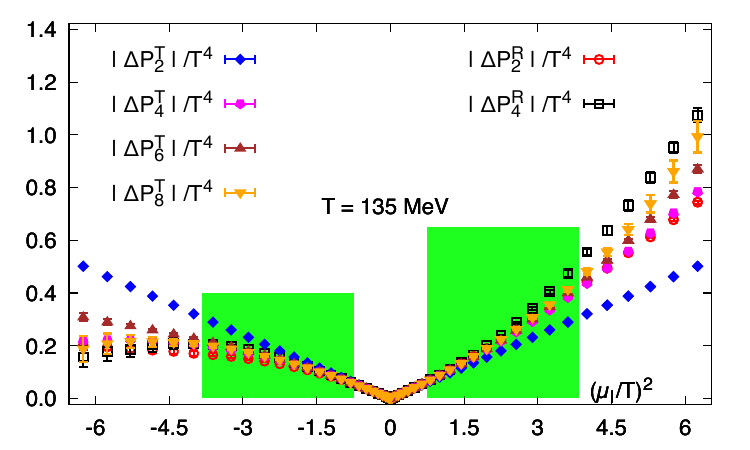} 
    \includegraphics[width=.32\textwidth]{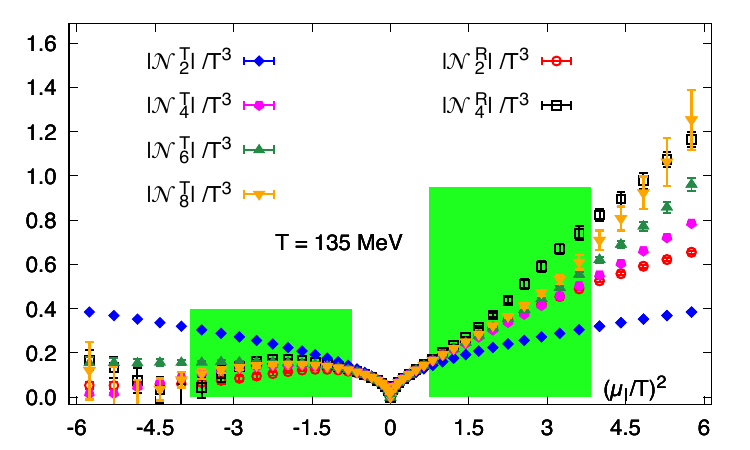}
    \includegraphics[width=.32\textwidth]{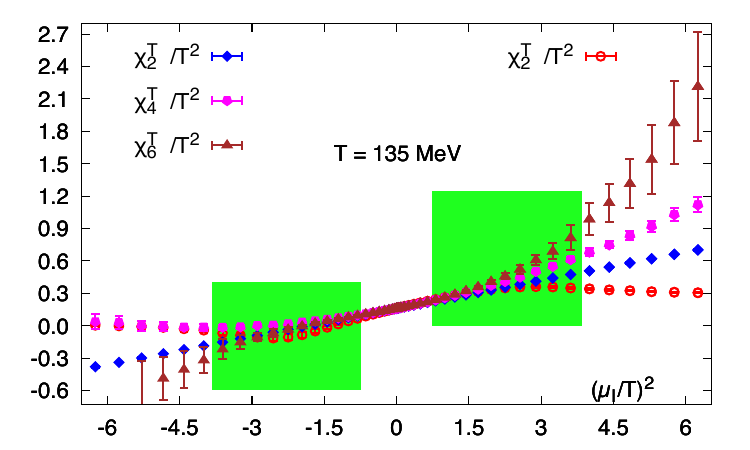} \\
    \includegraphics[width=.32\textwidth]{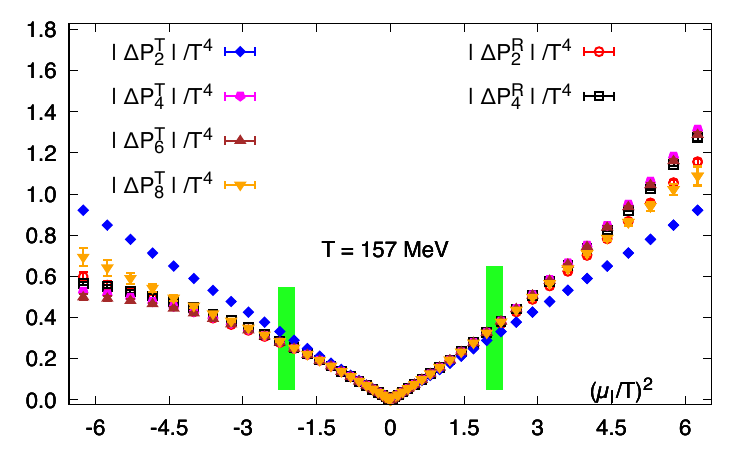} 
    \includegraphics[width=.32\textwidth]{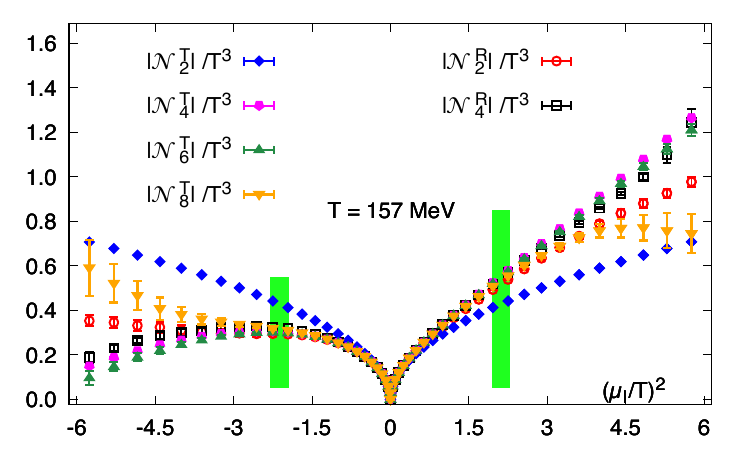} 
    \includegraphics[width=.32\textwidth]{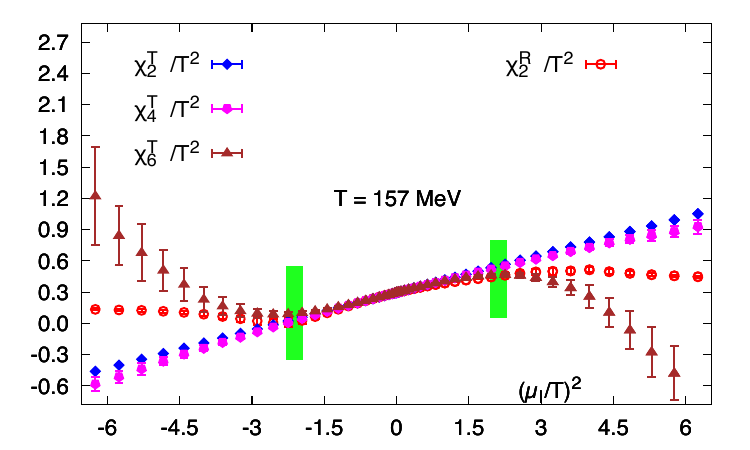} \\
    \includegraphics[width=.32\textwidth]{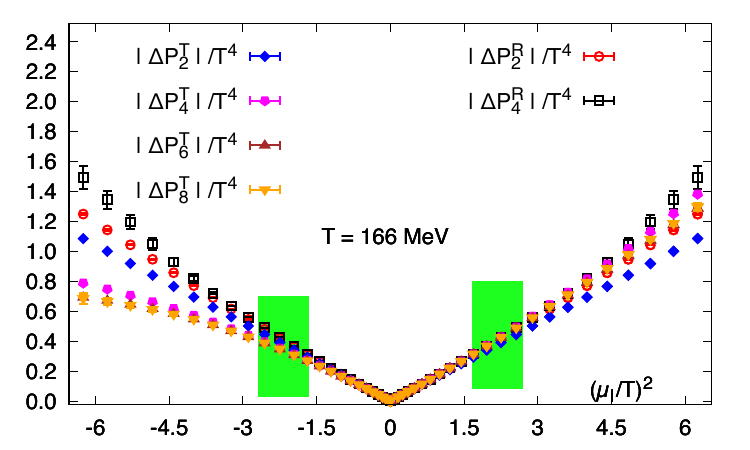}
    \includegraphics[width=.32\textwidth]{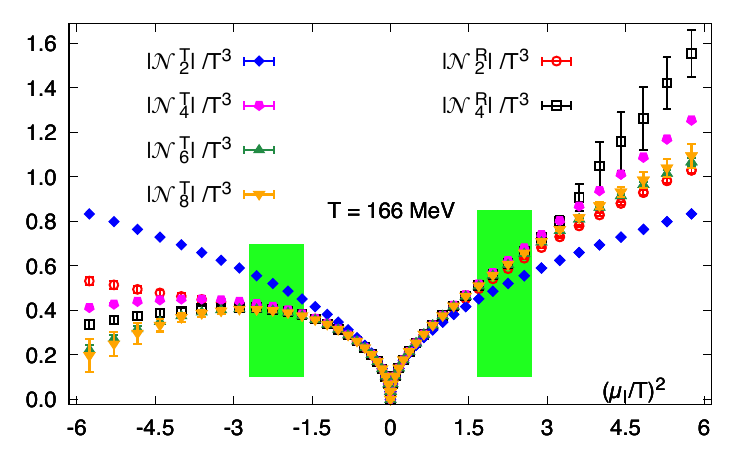}
    \includegraphics[width=.32\textwidth]{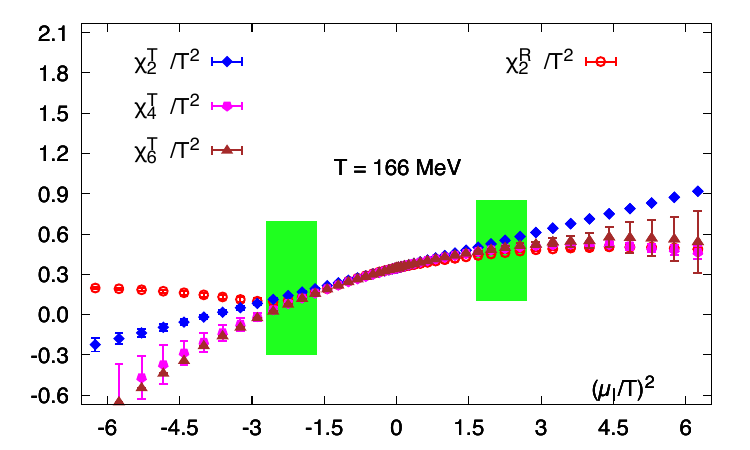} 
    \caption{The Taylor and unbiased Resummed results for the first three cumulants $\Delta P$ (left column), $\mathcal{N}$ (middle) and $\chi$ (right column) for real and imaginary values of $\muI$ for $T=135$ (top row), $157$ (middle) and $166$ MeV (bottom row) respectively. The green bands show the resummed estimate $\ROCresum$ in real and imaginary $\muI$ regimes.}
    \label{fig:Taylor and Resummed results for all with ROC}
    \end{minipage}
\end{figure*} 

%
%
  For these reasons, we compare the resummed results with only the different order Mercer-Roberts estimates $M_n$ for $n=2,4$ as given in Eqn.\eqref{eq:MR}. This comparison is illustrated in Fig.\,\ref{fig:MR}. For $M_4$\,, we also highlight the fourth order cumulant $\chi_2$ defined as 

\begin{equation*}
    \chi_2 = \frac{\partial^2 \left(\chi/T^2\right)}{\partial\, \left(\muI/T\right)^2}\Bigg|_{\muI=0} 
\end{equation*}
  The results are shown in Fig.\,\ref{fig:MR}. The more reliable $M_4$ out of the two estimates are shown by the red points in this figure. Like before, we observe here too that $M_2$ and $M_4$ reduce monotonically with increasing order of cumulants. The larger errorbars in $M_4$ as noticed are obvious, because of the prescence of less precise higher-order Taylor coefficients $c_6$ and $c_8$.  
  Nevertheless within errorbars, we find both $M_2$ and $M_4$ approach the resummed results with increasing order of the partition function cumulants. Because of some disagreement noticed at the level of third cumulant $\chi/T^2$, we construct a fourth one $\chi_2$ which has been defined before. With this new cumulant, we observe agreement to a better extent between $M_4$ and the resummed $\ROCresum$ estimate. Despite still having some discrepancies at $157$ and $166$ MeV, the overall agreement achieved within errors for all the other working temperatures in Fig.\,\ref{fig:MR} is an encouraging sight. This very much highlights the utility of adopting this resummation approach and thereby using it to determine Lee-Yang zeros and subsequent estimate of the radius of convergence, which we find can often well-indicate the same for the higher order Taylor series expansion of higher order cumulants of QCD partition function.

\subsection{Comparison among different order cumulants of the isospin partition function}
\label{subsec: comparing Cumulants}

We extend this comparative discussion to the level of individual cumulants by highlighting the Taylor and the unbiased exponential Resummed results of the first three cumulants. We present this comparison for three temperatures namely $135,157$ and $166$ MeV in Fig.\,\ref{fig:Taylor and Resummed results for all with ROC}, where we also demonstrate these results as function of real and imaginary $\muI$ values on the same plots. 

\begin{figure*}[ht]
\begin{minipage}[t]{.99\textwidth}
    \centering
    \includegraphics[width=.324\textwidth]{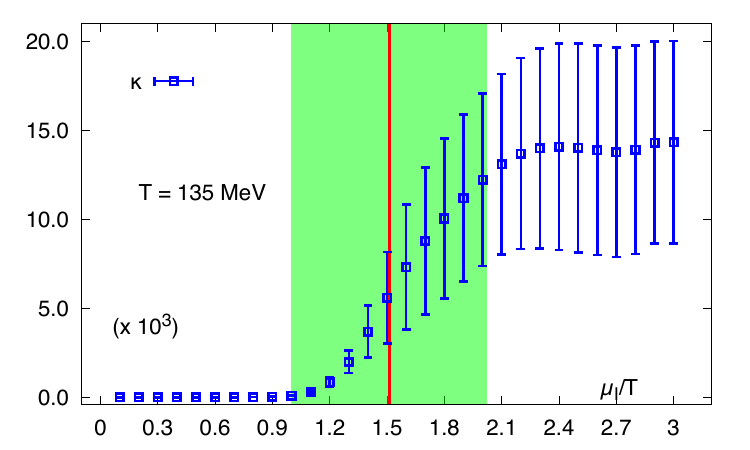}
    \includegraphics[width=.324\textwidth]{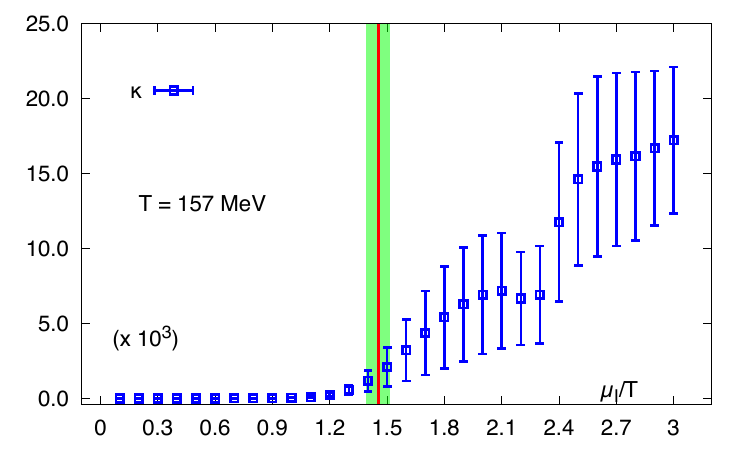}
    \includegraphics[width=.324\textwidth]{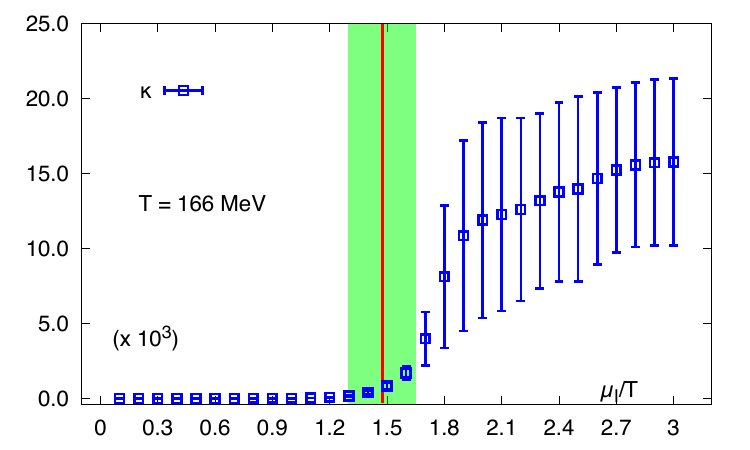}  
    \caption{Plots demonstrating kurtosis $\kappa$ as a function of real $\muI$ for $135$ (left),$157$ (middle) and $166$ MeV (right) respectively. The red line shows the mean value of $\ROCresum$, and the green band shows the same along with the associated errors.}    
    \label{fig:kurtosis} 
    \end{minipage}
\end{figure*}

We perform the unbiased exponential resummation series of these cumulants to second and fourth orders in real and imaginary $\muI$, whereas the familiar and already-published Taylor series results are extracted till the eighth orders. The resummed $\ROCresum$ estimate are depicted by the green bands in the figure. For all these three cumulants in Fig.\,\ref{fig:Taylor and Resummed results for all with ROC}, we clearly observe that the lower second order Taylor series results, shown by the blue points diverge away more quickly than the higher order results, for both real and imaginary $\muI$. As already proven in our previous paper \cite{Mitra2023unb_exp_res}, we observe and re-establish that the lower order resummed results agree appreciably with similar and higher order Taylor series results. However, the upshot and important observation in this case is this agreement sustains till encountering the green bands which depict the resummed estimate $\ROCresum$ in Fig.\,\ref{fig:Taylor and Resummed results for all with ROC}. Beyond this estimate \textit{i.e.} for $|\muI| > \ROCresum$, we find the different order calculations of resummed and Taylor series not only diverge away from one another but also importantly start manifesting loss of monotonic behaviour which otherwise as we observe, remains present for $|\muI| \leq \ROCresum$. In this regard, we also observe from Fig.\,\ref{fig:Taylor and Resummed results for all with ROC} that this order-by-order departure and lack of monotonicity become increasingly more distinct and seemingly more precise with increasing order of the cumulants.

\section{Overlap problem and Results} 
\label{sec:Overlap problem}

We provide a brief inspection on the overlap problem in this section. This problem and its degree of severity for finite $\muI$ simulations is important to understand the extent of reliability of using zero $\muI$ gauge ensembles in finite $\muI$ simulations.
While calculations for finite $\muI$ do not possess a sign problem, one always needs to exercise caution and be careful about the ever-present overlap problem. This is because, the degree of extrapolation to finite $\muI$ from $\muI=0$ loses its reliability and efficacy once there is no significant overlap between the distributions of the reweighting factors between the target ensemble at a given $\muI$ and the simulated known ensemble at zero $\muI$. Thus, this overlap becomes insignificant for larger values of $\muI$, leading to greater degree of severity of the overlap problem.

For a given gauge field configuration the extent of overlap is quantified by $\det \M(\muI)/\det \M(0)$, an unbiased estimate of which is given by

\begin{align}
    \frac{\det \M\left(\muI,U\right)}{\det \M\left(0,U\right)} &= \exp\Bigg[\nsum_{n=1}^N \left(\frac{\muI}{T}\right)^n \frac{\C_n\left(U\right)}{n!}\Bigg] 
    \label{eq:ratio for a single gauge configuration}
\end{align}
where $\C_n$ follow from Eqn.\eqref{eq:QCD exponential resummation}. A direct measurement of this problem can however be obtained from kurtosis $\kappa$ of the distribution of the determinant ratio, outlined in Eqn.\eqref{eq:ratio for a single gauge configuration}. This is the standardised fourth order central moment defined as

\begin{equation}
    \kappa = \frac{M_4\left(\muI\right)}{\big[\sigma\left(\muI\right)\big]^4}
    \label{eq:kurtosis}
\end{equation}  
where we have the fourth central moment $M_4$ and the standard deviation $\sigma$ as a function of $\muI$.

Fig.\,\ref{fig:kurtosis} portrays the behaviour of $\kappa$ as a function of $\muI$ for the three working temperatures $135,157$ and $166$ MeV. The red line, situated in the middle of the green band illustrates the mean value of $\ROCresum$. As expected, we very clearly observe that kurtosis $\kappa$ increases with increasing values of $\muI$. More importantly, the errorbars start increasing too in a monotonic manner with increasing $\muI$ and right across the green band, they exude a sharp sudden increment. We also sight non-monotonicity in $\kappa$ and these errorbars in $\modmuI>\ROCresum$ regime. This very clearly thus reflects the well-known fact that across $\ROCresum$ from which the Lee-Yang zeros start appearing in the complex $\muI$ plane, this extrapolation to finite $\muI$ from $\muI=0$ fails seriously and therefore working with a gauge ensemble simulated at $\muI=0$ is not viable.

\section{Conclusions and Outlook} 
\label{sec:Conclusions}

In this paper, we have thus demonstrated the Lee-Yang zeros of the QCD partition function in the complex isospin chemical potential $\muI$ plane, following the approach of unbiased exponential resummation. On course of this work, we have managed to determine a possible estimate (see Eqn.\eqref{eq:critical point}) of one critical point situated on the second order pion phase critical line in the low $T$, small $\muI$ regime of the QCD isospin phase diagram. As already outlined in this paper, an important feature of this method is the procedure of estimating these zeros directly from the path-integral form of the partition function itself. Unlike some of the previously related works in this direction, the Lee-Yang zeros and the subsequent $(T_I^c,\muI^c)$ are determined here entirely without utilizing any form of subsequent cumulants at any stage of the calculation, and also without invoking any property of this phase transition in the form of its order, universality class and point of onset. 

We have explicitly demonstrated in Fig.\ref{fig:roots of Z}, the various estimates of Lee-Yang zeros in the complex $\muI$ plane for a wide range of temperatures from $125$ to $171$ MeV including the crossover $T=157$ MeV. As mentioned before, this full estimation has been done utilizing the familiar Newton-Raphson method on the polynomial form of partition function, which we have obtained in an approximate form here by using the recent unbiased exponential resummation approach. For our work outlined here, the Lee-Yang zeros closest to origin in this complex $\muI$ plane are relevant, since they offer the estimate of the radius of convergence of the Taylor expansion of free energy at $\muI=0$. As it marks the upper bound of validity in $\muI$ starting from $\muI=0$ of possible Taylor expansions which entirely rely on extrapolations from $\muI=0$ or other $\muI < m_\pi$ values belonging in the uncondensed phase, this radius of convergence has been believed to indicate and outline possibly the second order pion condensate critical line in the isospin phase diagram at least for lower values of $T,\muI$. This critical line therefore outlines the upper limit of this uncondensed non-pion phase. For drawing meaningful results and implications from these nearest Lee-Yang zeros, we also have verified their stability in Sec.\ref{subsec:stability} by implementing the Newton-Raphson method on these very zero estimates, treating them as the new initial guesses. This is essential in realizing the reliability of these zeros which we have shown here explicitly, to be trustworthy and commendable in sense that the Newton-Raphson algorithm can identify these zeros successfully when using them as the corresponding initial guess values. 
This is what we have considered being reliable at least within the current precision and goal of this work and also given the relevant associated observables considered in this paper. This has been outlined thoroughly in Sec.\ref{subsec:stability} through Figs.\ref{fig:stability} and \ref{fig:Bastian-135}. 

In this connection, we have illustrated that while the imaginary parts of these zeros monotonically reduce with lowering $T$, the real parts do fluctuate but still manifest a sort of saturating feature and behaviour towards $\muI=m_\pi$ from $T \leq 135$ MeV considering also the associated errorbars. This bodes well in agreement with the conclusions of some previous works stating that from $T=140$ MeV, the pion phase boundary becomes vertical and without too much of a thermal variation, converges to $\muI=m_\pi$ at zero $T$. Although the radius of convergence in Fig.\ref{fig:ROC} of this work shows some discrepancy to this notion presumably because of finite imaginary part of these zeros, its behaviour showcasing a monotonic reduction with decreasing $T$ at least in the hadronic regime, is consistent with the chiral perturbation theory implications which hold true in this $\muI< m_\rho$ regime of exploration, where $m_\rho=770$ MeV. By means of a linear extrapolation which we perform solely in the spirit of following the thermal trend of the Lee-Yang zeros and nothing to do regarding the usual differentiation between crossover and phase transitions, we have determined the possible estimate of a critical point and we have demonstrated vividly in Sec.\ref{subsec:estimates from extrapolation of zeros for lower T} that this critical point is located at $\muI=m_\pi$ essentially within errorbars around $T=80$ MeV, which we show in Fig.\ref{fig:fitting} and observe that the imaginary part $\quickLYIm$ of the nearest Lee-Yang zero vanishes, consequently positioning it on the real $\muI$ axis. 
Even though this estimate of $T_I^c$ is less than $140$ MeV, the critical point thus obtained, converges to a true one on the second order critical line. The present study can be argued to provide a possible lower bound in the sense, that the second order pion condensate line can be predicted to exist at least for $T\in[0:T_I^c]$ within errorbars. Surely as mentioned before, several further studies are to be done involving finite-size scaling analysis of these zeros and their behaviour in thermodynamic limit, which are expected to provide a better picture outlining possible reasons of this difference, besides manifesting real zeros for all $T \leq 140$ MeV, and thereby aligning with the present state-of-the-art QCD isospin phase diagram at least in this regime. We consider all these crucial, important studies outside the scope of this current work and reserve them for interesting future works, with the belief that this will surely enlighten and establish further aspects of the phase diagram especially in the low $T$ regime, close to $T=0$ from this approach. We also plan to approach thermodynamic and continuum limits in the future, and subsequently investigate the location of these zeros, with lower values of tolerance for the Newton-Raphson method. Nevertheless, the very observation of the linearly extrapolated radius of convergence converging to a possible valid critical point on the second order pion phase boundary for $T=80$ MeV, is a novel feature of this work, given its present scope and limitations. 

We have also outlined in Sec.\ref{subsec: Ratio and MR estimates} of this paper, a detailed discussion of the resummed version $\ROCresum$ of the dimensionless radius of convergence by comparing it with the well-known Mercer-Roberts estimates of the same obtained via the Taylor series. These are illustrated by considering the series expansions of different order cumulants, starting from excess pressure (zeroth cumulant) to the fourth cumulant $\chi_2$.  We also have extended our analysis to include comparisons between Taylor and unbiased exponential results in real and imaginary $\muI$ values for the first three cumulants of the QCD partition function, which we have illustrated in Fig.\ref{fig:Taylor and Resummed results for all with ROC}. All these indicate that these results sustain order-by-order convergence within and upto $\ROCresum$, beyond which they show lack of monotonicity as well as divergences among different orders. Finally in the last section \ref{sec:Overlap problem} of the paper, we show that the overlap problem becomes severely drastic starting from this radius of convergence. Indicated by the stark increment of kurtosis with sudden rise in associated errorbars, this reflects minimal overlap between the target and simulated distributions starting from this regime and due to which, the zero $\muI$ extrapolation used, fails and lacks reliability here. Overall with  limitations and scope of further improvements, we have attempted to present in this paper, that unbiased exponential resummation approach can offer a reliable resummed estimate of the radius of convergence through Lee-Yang zero estimates and this has the potential to locate one of the constituent critical points on the critical second-order pion phase boundary line in the QCD isospin phase diagram. All data of our calculations can be found in Ref.\,\cite{Mitra2025data}.

\acknowledgments

 I sincerely acknowledge Jishnu Goswami for useful discussion and suggestions for this draft. I also thank all the other members of the HotQCD collaboration for their valuable inputs as well as for allowing me to use their data for the respective Taylor expansion calculations. The computations in this work have been performed on the GPU cluster at Bielefeld University, Germany. I also heartily thank the Bielefeld HPC.NRW team for their wholehearted support.

\bibliographystyle{plain}

\begin{thebibliography}{99}


\bibitem{Shuryak1978ij}
E.~V.~Shuryak,  
\textit{Quark-Gluon Plasma and Hadronic Production of Leptons, Photons and Psions},  
Phys.\ Lett.\ B \textbf{78}, 150 (1978)  
doi:10.1016/0370-2693(78)90370-2.


\bibitem{ALICE2008ngc}
K.~Aamodt \textit{et al.} [ALICE Collaboration],  
\textit{The ALICE experiment at the CERN LHC},  
JINST \textbf{3}, S08002 (2008)  
doi:10.1088/1748-0221/3/08/S08002.

\bibitem{Gyulassy2004RHIC}
M.~Gyulassy and L.~McLerran, 
\textit{New forms of QCD matter discovered at RHIC}, 
Nucl. Phys. A \textbf{750}, 30--63 (2005), 
doi:10.1016/j.nuclphysa.2004.10.034, 
arXiv:nucl-th/0405013.

\bibitem{PHENIX2004RHIC}
K. Adcox et al. (PHENIX Collaboration), 
\textit{Formation of dense partonic matter in relativistic nucleus-nucleus collisions at RHIC: Experimental evaluation by the PHENIX collaboration}, 
Nucl. Phys. A \textbf{757}, 184--283 (2005), 
doi:10.1016/j.nuclphysa.2005.03.086.

\bibitem{Aad2008ATLAS}
G. Aad et al., 
\textit{The ATLAS Experiment at the CERN Large Hadron Collider}, 
JINST \textbf{3}, S08003 (2008), 
doi:10.1088/1748-0221/3/08/S08003.



\bibitem{Busza2018RHIC}
W.~Busza, K.~Rajagopal, and W.~V.~Schee, 
\textit{Heavy Ion Collisions: The Big Picture, and the Big Questions}, 
Ann. Rev. Nucl. Part. Sci. \textbf{68}, 339--376 (2018), 
doi:10.1146/annurev-nucl-101917-020852, 
arXiv:1802.04801.

\bibitem{Ratti2018RHIC}
C.~Ratti, 
\textit{Lattice QCD and heavy ion collisions: a review of recent progress}, 
Rept. Prog. Phys. \textbf{81}, 084301 (2018), 
doi:10.1088/1361-6633/aabb97, 
arXiv:1804.07810.


\bibitem{Baym1971star}
G.~Baym, H.~A.~Bethe, and C.~Pethick, 
\textit{Neutron star matter}, 
Nucl. Phys. A \textbf{175}, 225--271 (1971), 
doi:10.1016/0375-9474(71)90281-8.

\bibitem{Shuryak1980star}
E.~V.~Shuryak, 
\textit{Quantum Chromodynamics and the Theory of Superdense Matter}, 
Phys. Rept. \textbf{61}, 71--158 (1980), 
doi:10.1016/0370-1573(80)90105-2.

\bibitem{Baym2017star}
G.~Baym, T.~Hatsuda, T.~Kojo, P.~D.~Powell, Y.~Song, and T.~Takatsuka, 
\textit{From hadrons to quarks in neutron stars: a review}, 
Rept. Prog. Phys. \textbf{81}, 056902 (2018), 
doi:10.1088/1361-6633/aaae14, 
arXiv:1707.04966.

\bibitem{Barshay1973pioncond}
S. Barshay, G. Vagradov, and G. E. Brown, 
\textit{Possibility of a phase transition to a pion condensate in neutron stars}, 
Phys. Lett. B \textbf{43}, 359--361 (1973), 
doi:10.1016/0370-2693(73)90370-5.

\bibitem{Kolb1990uni}
E.~W.~Kolb, 
\textit{The Early Universe}, 
Taylor and Francis, 2019, 
Vol. 69, 
ISBN: 978-0-429-49286-0, 978-0-201-62674-2, 
doi:10.1201/9780429492860.

\bibitem{Halasz1998qcd_pha_dia}
Adam Miklos Halasz, A. D. Jackson, R. E. Shrock, Misha A. Stephanov, and J. J. M. Verbaarschot, 
\textit{On the phase diagram of QCD}, 
Phys. Rev. D \textbf{58}, 096007 (1998), 
doi:10.1103/PhysRevD.58.096007, 
arXiv:hep-ph/9804290.


\bibitem{Karsch2001QCD}
F. Karsch, 
\textit{Lattice QCD at high temperature and density}, 
in \textit{Lect. Notes Phys.}, vol. 583, 
209--249 (2002), 
doi:10.1007/3-540-45792-5\_6, 
arXiv:hep-lat/0106019.

\bibitem{HotQCD2014QCD}
A. Bazavov et al. (HotQCD), 
\textit{Equation of state in (2+1)-flavor QCD}, 
Phys. Rev. D \textbf{90}, 094503 (2014), 
doi:10.1103/PhysRevD.90.094503, 
arXiv:1407.6387.


\bibitem{HotQCD2018crossover}
A. Bazavov et al. (HotQCD), 
\textit{Chiral crossover in QCD at zero and non-zero chemical potentials}, 
Phys. Lett. B \textbf{795}, 15--21 (2019), 
doi:10.1016/j.physletb.2019.05.013, 
arXiv:1812.08235.

\bibitem{Steinbrecher2018crossover}
P. Steinbrecher, 
\textit{The QCD crossover at zero and non-zero baryon densities from Lattice QCD}, 
Nucl. Phys. A \textbf{982}, 847--850 (2019), 
doi:10.1016/j.nuclphysa.2018.08.025, 
arXiv:1807.05607.


\bibitem{Borsanyi2020crossover}
S. Borsanyi et al., 
\textit{QCD Crossover at Finite Chemical Potential from Lattice Simulations}, 
Phys. Rev. Lett. \textbf{125}, 052001 (2020), 
doi:10.1103/PhysRevLett.125.052001, 
arXiv:2002.02821.

\bibitem{Karsch1988fin_den_qcd}
F. Karsch and K. H. Mutter, 
\textit{Strong Coupling QCD at finite Baryon Number Density}, 
Nucl. Phys. B \textbf{313}, 541--559 (1989), 
doi:10.1016/0550-3213(89)90396-9, 
CERN-TH-5063/88.

\bibitem{Karsch1999fin_den_qcd}
F. Karsch, 
\textit{Lattice QCD at finite temperature and density}, 
Nucl. Phys. B Proc. Suppl. \textbf{83}, 14--23 (2000), 
doi:10.1016/S0920-5632(00)91591-3, 
arXiv:hep-lat/9909006.

\bibitem{deForcrand2009fin_den_qcd}
P. Forcrand, 
\textit{Simulating QCD at finite density}, 
PoS LAT2009, 010 (2009), 
doi:10.22323/1.091.0010, 
arXiv:1005.0539.

\bibitem{Fu2019fin_den_qcd}
W.J. Fu, J. M. Pawlowski, and F. Rennecke, 
\textit{QCD phase structure at finite temperature and density}, 
Phys. Rev. D \textbf{101}, 054032 (2020), 
doi:10.1103/PhysRevD.101.054032, 
arXiv:1909.02991.



\bibitem{Fodor2001crit}
Z.~Fodor and S.~D.~Katz, 
\textit{Lattice determination of the critical point of QCD at finite T and mu}, 
JHEP \textbf{03}, 014 (2002), 
doi:10.1088/1126-6708/2002/03/014, 
arXiv:hep-lat/0106002.

\bibitem{Karsch2001crit}
F.~Karsch, E.~Laermann, and C.~Schmidt, 
\textit{The Chiral critical point in three-flavor QCD}, 
Phys. Lett. B \textbf{520}, 41--49 (2001), 
doi:10.1016/S0370-2693(01)01114-5, 
arXiv:hep-lat/0107020.

\bibitem{Gavai2004crit}
R.~V.~Gavai and S.~Gupta, 
\textit{The Critical end point of QCD}, 
Phys. Rev. D \textbf{71}, 114014 (2005), 
doi:10.1103/PhysRevD.71.114014, 
arXiv:hep-lat/0412035.



\bibitem{Barbour1986jf}
I. Barbour, N.E. Behilil, E. Dagotto, F. Karsch, A. Moreo, M. Stone, and H. W. Wyld, 
\textit{Problems with Finite Density Simulations of Lattice QCD}, 
Nucl. Phys. B \textbf{275}, 296--318 (1986), 
doi:10.1016/0550-3213(86)90601-2.

\bibitem{Fodor2001au}
Z.~Fodor and S.~D.~Katz,
\textit{A New method to study lattice QCD at finite temperature and chemical potential},
Phys.\ Lett.\ B \textbf{534}, 87--92 (2002),
doi:10.1016/S0370-2693(02)01583-6
[arXiv:hep-lat/0104001].


\bibitem{Pan2022fgf}
G. Pan and Z.Y. Meng, 
\textit{Sign Problem in Quantum Monte Carlo Simulation}, 
arXiv:2204.08777 (2022), 
doi:10.1016/B978-0-323-90800-9.00095-0.


\bibitem{Gavai2004Taylor}
R. V. Gavai and S. Gupta, 
\textit{The Critical end point of QCD}, 
Phys. Rev. D \textbf{71}, 114014 (2005), 
doi:10.1103/PhysRevD.71.114014, 
arXiv:hep-lat/0412035.

\bibitem{deForcrand2003IMag}
P. de Forcrand and O. Philipsen, 
\textit{QCD phase diagram at small densities from simulations with imaginary $\mu$}, 
in \textit{5th International Conference on Strong and Electroweak Matter}, 271--275 (2003), 
doi:10.1142/9789812704498\_0027, 
arXiv:hep-ph/0301209.


\bibitem{Alexandru2015thimbles}
A. Alexandru, G. Basar, and P. Bedaque, 
\textit{Monte Carlo algorithm for simulating fermions on Lefschetz thimbles}, 
Phys. Rev. D \textbf{93}, 014504 (2016), 
doi:10.1103/PhysRevD.93.014504, 
arXiv:1510.03258.


\bibitem{Sexty2013ComLange}
D. Sexty, 
\textit{Simulating full QCD at nonzero density using the complex Langevin equation}, 
Phys. Lett. B \textbf{729}, 108--111 (2014), 
doi:10.1016/j.physletb.2014.01.019, 
arXiv:1307.7748.


\bibitem{Mondal2021exp_res}
S. Mondal, S. Mukherjee, and P. Hegde, 
\textit{Lattice QCD Equation of State for Nonvanishing Chemical Potential by Resumming Taylor Expansions}, 
Phys. Rev. Lett. \textbf{128}, 022001 (2022), 
doi:10.1103/PhysRevLett.128.022001, 
arXiv:2106.03165.

\bibitem{Bollweg2022Pade}
D. Bollweg et al., 
\textit{Taylor expansions and Pad\'e approximants for cumulants of conserved charge fluctuations at nonvanishing chemical potentials}, 
Phys. Rev. D \textbf{105}, 074511 (2022), 
doi:10.1103/PhysRevD.105.074511, 
arXiv:2202.09184.

\bibitem{Mitra2023unb_exp_res}
S. Mitra and P. Hegde, 
\textit{QCD equation of state at finite chemical potential from an unbiased exponential resummation of the lattice QCD Taylor series}, 
Phys. Rev. D \textbf{108}, 034502 (2023), 
doi:10.1103/PhysRevD.108.034502, 
arXiv:2302.06460.


\bibitem{Son2000xc}
D.~T.~Son and M.~A.~Stephanov,  
\textit{QCD at finite isospin density},  
Phys.\ Rev.\ Lett.\ \textbf{86}, 592--595 (2001)  
doi:10.1103/PhysRevLett.86.592  
[arXiv:hep-ph/0005225].


\bibitem{deForcrand2007isophdia}
P. Forcrand, M. A. Stephanov, and U. Wenger, 
\textit{On the phase diagram of QCD at finite isospin density}, 
PoS LATTICE2007, 237 (2007), 
arXiv:0711.0023.

\bibitem{Andersen2023ofv}
J.~O.~Andersen, M.~Johnsrud, Q.~Yu and H.~Zhou,  
\textit{Chiral perturbation theory and Bose-Einstein condensation in QCD},  
Phys.\ Rev.\ D \textbf{111}, Vol. 3, 034017 (2025),  
doi:10.1103/PhysRevD.111.034017,  
arXiv:2312.13092 [hep-ph].

\bibitem{Brandt2017isophdia}
B. B. Brandt, G. Endrodi, and S. Schmalzbauer, 
\textit{QCD phase diagram for nonzero isospin-asymmetry}, 
Phys. Rev. D \textbf{97}, 054514 (2018), 
doi:10.1103/PhysRevD.97.054514, 
arXiv:1712.08190.


\bibitem{Brandt2019isophdia}
B. B. Brandt, F. Cuteri, G. Endrodi, and S. Schmalzbauer, 
\textit{Exploring the QCD phase diagram via reweighting from isospin chemical potential}, 
PoS \textbf{LATTICE2019}, 189 (2019),
doi:10.22323/1.363.0189
arXiv:1911.12197.


\bibitem{Yokota2023osv}
T.~Yokota et. al.,  
\textit{Color superconductivity on the lattice — analytic predictions from QCD in a small box},  
JHEP \textbf{06}, 061 (2023)  
doi:10.1007/JHEP06(2023)061  
[arXiv:2302.11273 [hep-lat]].


\bibitem{Reinosa2015oua}
U.~Reinosa, J.~Serreau and M.~Tissier,  
\textit{Perturbative study of the QCD phase diagram for heavy quarks at nonzero chemical potential},  
Phys.\ Rev.\ D \textbf{92}, 025021 (2015)  
doi:10.1103/PhysRevD.92.025021  
[arXiv:1504.02916 [hep-th]].



\bibitem{Scherer2002tk}
S.~Scherer,  
\textit{Introduction to chiral perturbation theory},  
Adv.\ Nucl.\ Phys.\ \textbf{27}, 277 (2003),  
[arXiv:hep-ph/0210398].

\bibitem{Alexandru2005la}
A.~Alexandru, M.~Faber, I.~Horv\'ath and K.~F.~Liu,  
\textit{Lattice QCD at finite density via a new canonical approach},  
Phys.\ Rev.\ D \textbf{72}, 114513 (2005)  
doi:10.1103/PhysRevD.72.114513,  
arXiv:hep-lat/0507020.

\bibitem{Mitra2023thesis}
S.~Mitra, 
\textit{Exponential resummation of QCD at finite chemical potential}, 
PhD thesis, Indian Institute of Science Bangalore, Centre for High Energy Physics, India, 2023, 
arXiv:2307.05751.


\bibitem{Dimopoulos2021vrk}
P.~Dimopoulos et. al.,  
\textit{Contribution to understanding the phase structure of strong interaction matter: Lee-Yang edge singularities from lattice QCD},  
Phys.\ Rev.\ D \textbf{105}, no.3, 034513 (2022)  
doi:10.1103/PhysRevD.105.034513  
[arXiv:2110.15933 [hep-lat]].

\bibitem{Schmidt2022ogw}
C.~Schmidt et. al.,  
\textit{Detecting Critical Points from the Lee--Yang Edge Singularities in Lattice QCD},  
Acta Phys.\ Polon.\ Supp.\ \textbf{16}, no.1, 1--A52 (2023)  
doi:10.5506/APhysPolBSupp.16.1-A52  
[arXiv:2209.04345 [hep-lat]].

\bibitem{Clarke2024seq}
D.~A.~Clarke et. al.,  
\textit{Searching for the QCD critical point using Lee-Yang edge singularities},  
PoS \textbf{LATTICE2023}, 168 (2024)  
doi:10.22323/1.453.0168  
[arXiv:2401.08820 [hep-lat]].


\bibitem{Clarke2024ugt}
D.~A.~Clarke et. al.,  
\textit{Searching for the QCD critical endpoint using multi-point Pad\'e approximations},  
[arXiv:2405.10196 [hep-lat]].


\bibitem{LeeYang1952}
T. D. Lee and C. N. Yang, 
\textit{Statistical Theory of Equations of State and Phase Transitions. II. Lattice Gas and Ising Model}, 
Phys. Rev. \textbf{87}, 410--419 (1952), 
doi:10.1103/PhysRev.87.410.







\bibitem{Mercer1990}
G. N. Mercer and A. J. Roberts, 
\textit{A Centre Manifold Description of Contaminant Dispersion in Channels with Varying Flow Properties}, 
SIAM Journal on Applied Mathematics \textbf{50}, 1547--1565 (1990), 
doi:10.1137/0150091, 
\url{https://doi.org/10.1137/0150091}.


\bibitem{Vovchenko2017cluster}
Volodymyr Vovchenko, Jan Steinheimer, Owe Philipsen, and Horst Stoecker, 
\textit{Cluster Expansion Model for QCD Baryon Number Fluctuations: No Phase Transition at $\mu_B / T < \pi$}, 
Phys. Rev. D \textbf{97}, 114030 (2018), 
doi:10.1103/PhysRevD.97.114030, 
arXiv:1711.01261.

\bibitem{Pasztor2019mercer}
Matteo Giordano and Attila P\'asztor, 
\textit{Reliable estimation of the radius of convergence in finite density QCD}, 
Phys. Rev. D \textbf{99}, 114510 (2019), 
doi:10.1103/PhysRevD.99.114510, 
arXiv:1904.01974.



\bibitem{Gavai2003tay}
R.~Gavai, S.~Gupta, and R.~Ray, 
\textit{Taylor expansions in chemical potential}, 
Prog. Theor. Phys. Suppl. \textbf{153}, 270--276 (2004), 
doi:10.1143/PTPS.153.270, 
arXiv:nucl-th/0312010.

\bibitem{Ejiri2003tay}
S.~Ejiri et. al., 
\textit{Study of QCD thermodynamics at finite density by Taylor expansion}, 
Prog. Theor. Phys. Suppl. \textbf{153}, 118--126 (2004), 
doi:10.1143/PTPS.153.118, 
arXiv:hep-lat/0312006.

\bibitem{Miao2008tay}
C.~Miao and C.~Schmidt, 
\textit{Non-zero density QCD by the Taylor expansion method: The Isentropic equation of state, hadronic fluctuations and more}, 
PoS \textbf{LATTICE2008}, 172 (2008), 
doi:10.22323/1.066.0172, 
arXiv:0810.0375.

\bibitem{Karsch2010tay}
F. Karsch, B.-J. Schaefer, M. Wagner, and J. Wambach, 
\textit{Towards finite density QCD with Taylor expansions}, 
Phys. Lett. B \textbf{698}, 256--264 (2011), 
doi:10.1016/j.physletb.2011.03.013, 
arXiv:1009.5211.


\bibitem{Bollweg2022coefficients}
D. Bollweg, J. Goswami, O. Kaczmarek, F. Karsch, Swagato Mukherjee, P. Petreczky, C. Schmidt, and P. Scior, 
\textit{Taylor expansions and Pad\'e approximants for cumulants of conserved charge fluctuations at nonvanishing chemical potentials}, 
Phys. Rev. D \textbf{105}, 074511 (2022), 
doi:10.1103/PhysRevD.105.074511, 
arXiv:2202.09184.


\bibitem{Symanzik1983gau}
K. Symanzik, 
\textit{Continuum Limit and Improved Action in Lattice Theories. 1. Principles and $\varphi^4$ Theory}, 
Nucl. Phys. B \textbf{226}, 187--204 (1983), 
doi:10.1016/0550-3213(83)90468-6.

\bibitem{Follana2006rc}
E. Follana et al. (HPQCD, UKQCD), 
\textit{Highly improved staggered quarks on the lattice, with applications to charm physics}, 
Phys. Rev. D \textbf{75}, 054502 (2007), 
doi:10.1103/PhysRevD.75.054502, 
arXiv:hep-lat/0610092.


\bibitem{Bazavov2010hisq}
A. Bazavov and P. Petreczky, 
\textit{Deconfinement and chiral transition with the highly improved staggered quark (HISQ) action}, 
J. Phys. Conf. Ser. \textbf{230}, 012014 (2010), 
doi:10.1088/1742-6596/230/1/012014, 
arXiv:1005.1131.

\bibitem{Bazavov2011nk}
A. Bazavov et al., 
\textit{The chiral and deconfinement aspects of the QCD transition}, 
Phys. Rev. D \textbf{85}, 054503 (2012), 
doi:10.1103/PhysRevD.85.054503, 
arXiv:1111.1710.

\bibitem{Mitra2022cum_exp}
S.~Mitra, P.~Hegde, and C.~Schmidt, 
\textit{New way to resum the lattice QCD Taylor series equation of state at finite chemical potential}, 
Phys. Rev. D \textbf{106}, 034504 (2022), 
doi:10.1103/PhysRevD.106.034504, 
arXiv:2205.08517.

\bibitem{Mitra2022unb_exp_res}
S.~Mitra, 
\textit{Determination of Lattice QCD Equation of State at a Finite Chemical Potential}, 
Springer Proc. Phys. \textbf{304}, 209--212 (2024), 
doi:10.1007/978-981-97-0289-3\_45, 
arXiv:2209.11937.

\bibitem{Mitra2022Bonn_unb_exp_res}
S.~Mitra, P.~Hegde, and C.~Schmidt, 
\textit{A new way to resum Lattice QCD equation of state at finite chemical potential}, 
PoS \textbf{LATTICE2022}, 153 (2023), 
doi:10.22323/1.430.0153, 
arXiv:2209.07241.

\bibitem{NewtonRaphson}
T.~J.~Ypma, 
\textit{Historical Development of the Newton–Raphson Method}, 
SIAM Review \textbf{37}, 531--551 (1995), 
doi:10.1137/1037125, 
\url{https://doi.org/10.1137/1037125}.



\bibitem{Bazavov:2017dus}
A. Bazavov et al., 
\textit{The QCD Equation of State to $\mathcal{O}(\mu_B^6)$ from Lattice QCD}, 
Phys. Rev. D \textbf{95}, 054504 (2017), 
doi:10.1103/PhysRevD.95.054504, 
arXiv:1701.04325.

\bibitem{HotQCD:2018pds}
A. Bazavov et al., 
\textit{Chiral crossover in QCD at zero and non-zero chemical potentials}, 
Phys. Lett. B \textbf{795}, 15--21 (2019), 
doi:10.1016/j.physletb.2019.05.013, 
arXiv:1812.08235.

\bibitem{Bollweg:2021vqf}
D. Bollweg et al., 
\textit{Second order cumulants of conserved charge fluctuations revisited: Vanishing chemical potentials}, 
Phys. Rev. D \textbf{104}, 074512 (2021), 
doi:10.1103/PhysRevD.104.074512, 
arXiv:2107.10011.

\bibitem{Bollweg:2022rps}
D. Bollweg et al., 
\textit{Taylor expansions and Pad\'e approximants for cumulants of conserved charge fluctuations at nonvanishing chemical potentials}, 
Phys. Rev. D \textbf{105}, 074511 (2022), 
doi:10.1103/PhysRevD.105.074511, 
arXiv:2202.09184.

\bibitem{Bollweg:2022fqq}
D. Bollweg et al., 
\textit{Equation of state and speed of sound of (2+1)-flavor QCD in strangeness-neutral matter at nonvanishing net baryon-number density}, 
Phys. Rev. D \textbf{108}, 014510 (2023), 
doi:10.1103/PhysRevD.108.014510, 
arXiv:2212.09043.




\bibitem{Fodor2011lcp}
Z. Fodor and S. D. Katz, 
\textit{Lattice QCD thermodynamics}, 
Acta Phys. Polon. B \textbf{42}, 2791--2810 (2011), 
doi:10.5506/APhysPolB.42.2791.


\bibitem{Borsanyi2023tdp}
S.~Borsanyi, Z.~Fodor, M.~Giordano, J.~N.~Guenther, S.~D.~Katz, A.~Pasztor and C.~H.~Wong,
\textit{Can rooted staggered fermions describe nonzero baryon density at low temperatures?},
\emph{Phys. Rev. D} \textbf{109}, no.5, 054509 (2024),
doi:10.1103/PhysRevD.109.054509,
arXiv:2308.06105 [hep-lat].


\bibitem{Andersen2023ivj}
J.~O.~Andersen, Q.~Yu and H.~Zhou,
\textit{Pion condensation in QCD at finite isospin density, the dilute Bose gas, and speedy Goldstone bosons},
Phys.\ Rev.\ D \textbf{109}, no.3, 034022 (2024)
doi:10.1103/PhysRevD.109.034022
[arXiv:2306.14472 [hep-ph]].


\bibitem{Schmidt2024uuk}
C.~Schmidt,  
\textit{Lee-Yang edge singularities in QCD: From Fourier coefficients to parametrizations of the universal scaling functions},  
PoS \textbf{EuroPLEx2023} (2024) 037, doi:10.22323/1.451.0037.



\bibitem{Schmidt2025ftp}
C.~Schmidt, \textit{The QCD phase diagram, universal scaling, and Lee-Yang zeros}, arXiv:2501.19336 [hep-lat], Jan. 2025.

\bibitem{Mitra2025data}
S.~Mitra, 
\textit{Dataset for ``Estimates of Lee-Yang zeros and a possible critical point on the pion condensate boundary in the QCD isospin phase diagram using an unbiased exponential resummation on the lattice". Bielefeld University.}, https://doi.org/10.4119/unibi/3004966 



\end{thebibliography}

\onecolumngrid
\appendix

\newpage

\section{Fixing our choice of the initial convention of isospin}
\label{appendix sec: fixing the convention}

For setting up the relations between the chemical potentials of $u,d,s$ basis and our working $B,S,I$ basis in this work, we have considered the usual quantum number relations in (2+1)-flavor QCD as follows : 

 \begin{equation}
     B=\frac{1}{3}\left(N_u+N_d\right) \hspace{.5cm},\hspace{.5cm}  S=-N_s\hspace{.5cm}, \hspace{.5cm}    I = \frac{1}{2} \left(N_u-N_d\right)
     \label{eq:quantum numbers}
 \end{equation}
Here $B,S,I$ are baryon, strangeness and isospin quantum numbers respectively with $N_u,N_d,N_s$ being the number of up, down and strange quarks. 
 
 The isospin prefactor $1/2$ in the above Eqn.\eqref{eq:quantum numbers} comes from the fact that protons and neutrons have isospin $1/2$, since one can be formed from the other by interchanging an up quark with a down quark and vice-versa. Also, since they are baryons and therefore three-quark spin $1/2$ systems, the constituent up and down quarks should also have isospin quantum numbers $1/2$. The assignment of $+1/2$ to proton and $u$ quarks, while $-1/2$ to neutrons and $d$ quarks depends on one's choice. This motivates us to choose and assign in this work, each $u$ quark to have $I_u=1/2$ and each $d$ quark to have $I_d=-1/2$. Since, the property of strangeness comes entirely from the strange quark, we assign $I_s=-1$ to strange quark. This logically leads us to above Eqn.\eqref{eq:quantum numbers}. With this, naturally the pion quantum number $I_\pi=1$, as it is a two quark system (meson) and it comes in three species $\pi^+,\pi^-,\pi^0$. This also can be justified using the familiar quantum-mechanical relation $N_I=2\,I+1$. In this case, $I=1$ for pion and hence, the number of possible pions $N_I=3$. 
 
 With this set of equations in \eqref{eq:quantum numbers} and also the fugacity conservation equation $\muB\,B\,+\mu_S\,S\,+\muI\,I=\mu_u\,N_u\,+\mu_d\,N_d\,+\mu_s\,N_s$, we therefore find
\begin{equation}
    \muB=\frac{3}{2}\left(\mu_u+\mu_d\right) \hspace{.5cm}, \hspace{.4cm} \muI=\left(\mu_u-\mu_d\right).
\end{equation}
where $\muB,\muI$ are the baryon and isospin chemical potentials and $\mu_u,\mu_d$ are the same for up and down quarks respectively. This is precisely the starting convention used in this work, and also in a recent work \cite{Andersen2023ivj}.

\end{document}